\documentclass[pra,twocolumn,nofootinbib]{revtex4}
\usepackage{amsmath}
\usepackage{amssymb}
\usepackage{graphicx}
\usepackage{amscd}
\newcommand{\goesto}[1][]{\ensuremath{\begin{CD}@>{#1}>>\end{CD}}}
\begin{document}

\title{Quantum coherence and control in one- and two-photon optical systems}

\author{Andrew J. Berglund}
 \altaffiliation[Present
address: ]{Norman Bridge Laboratory of Physics 12-33, California
Institute of Technology, Pasadena, CA 91125,
USA}\email{berglund@its.caltech.edu}
 \affiliation{Department of
Physics and Astronomy, Dartmouth College, Hanover, NH 03755, USA}
 \affiliation{Physics Division, P-23, Los Alamos National
Laboratory, Los Alamos, NM 87545, USA}



\begin{abstract}
\par We investigate coherence in one- and two-photon optical systems, both
theoretically and experimentally. In the first case, we develop
the density operator representing a single photon state subjected
to a non-dissipative coupling between observed (polarization) and
unobserved (frequency) degrees of freedom. We show that an
implementation of ``bang-bang'' quantum control protects photon
polarization information from certain types of decoherence. In the
second case, we investigate the existence of a ``decoherence-free"
subspace of the Hilbert space of two-photon polarization states
under the action of a similar coupling. The density operator
representation is developed analytically and solutions are
obtained numerically.
\end{abstract}
\maketitle
\par [{\bf Note}: This manuscript is taken from the author's
undergraduate thesis (A.B. Dartmouth College, June 2000, advised
by Dr. Walter E. Lawrence), an experimental and theoretical
investigation under the supervision of Dr. Paul G.
Kwiat.\footnote{Physics Division, P-23, Los Alamos National
Laboratory, Los Alamos, NM 87545, USA. Email: kwiat@lanl.gov.}]
\section{Introduction}\label{intro}

    Decoherence in two-state quantum systems is a significant
obstacle to the realization of proposed quantum information
technologies. Coupling between quantum bit (``qubit'') states and
unobserved environmental degrees of freedom leads to decoherence
effects which limit the practical implementation of proposed
quantum algorithms \cite{unruh}. Photon modes, including
polarization and spatial modes, provide an easily accessible
system in which simple quantum circuits can be investigated
\cite{unitary,opticalqubits,grover}. Here, we examine the process
of decoherence by subjecting single photons to a controllable
birefringent ``environment'' and observing the evolution of the
polarization state.

    In section \ref{one-photon}, we investigate the evolution of a single photon
state under the action of a unitary coupling between polarization
and frequency modes.  Such non-dissipative ``phase errors'' give
rise to decoherence effects, whereby the photon evolves from a
definite polarization state to an unpolarized state. We then
introduce and examine an optical implementation of so-called
``bang-bang" quantum control of decoherence by rapidly exchanging
the eigenstates of the coupling operation \cite{bang-bang}. The
term ``quantum control" is justified since such an operation will
be shown to preserve a coherent polarization state in some special
cases, and to reduce decoherence in more general cases.

    In section \ref{DFS} we will
introduce a particular two-photon polarization-entangled state
that, due to its symmetry properties, is immune to collective
decoherence of the type mentioned above. That is, this state is a
decoherence-free subspace (DFS) of the Hilbert space of photon
polarization \cite{zanardi,lidar}. Photon pairs entangled in both
polarization and frequency degrees of freedom, such as
hyper-entangled photons produced in down-conversion sources (see
\cite{source,ultrabright}), further complicate this particular
decoherence mechanism . In particular, energy conservation imposes
frequency correlations which affect the coherence properties of
these two-photon states. In the experimental case, it will be
shown that this frequency correlation can be effectively
suppressed and the DFS recovered by a simple and physically
intuitive modification of the apparatus.

\section{Coherence and ``bang-bang" control: The one-photon
case}\label{one-photon}

\subsection{Review of single-photon decoherence}\label{review}

In this section, we will show that non-dissipative (unitary)
coupling between photon frequency and polarization in a
birefringent ``environment'' followed by a trace over frequency
leads to decoherence.\footnote{This effect has a well-known
counterpart in classical optics whereby quasi-monochromatic light
composed of uncorrelated frequency components loses the ability to
interfere with itself when polarization modes are separated beyond
the coherence length of the incident light, as in an unbalanced
Michelson interferometer (see \cite{born_and_wolf}, $\S$7.5.8).}
These results are relevant to the study of decoherence in quantum
systems which arises when coupling to environmental states,
followed by a trace over those degrees of freedom, leads to a loss
of phase information between qubit basis states. In the following
argument, the term ``environmental'' will be used to describe the
coupling of information-carrying states (photon polarization
modes) to degrees of freedom which are not utilized for
information representation (photon frequency modes). It will be
shown that even non-dissipative coupling between these modes leads
to a loss of information in the qubit states.

A single photon characterized by its frequency spectrum and
polarization can be represented by the state ket
\begin{equation} |\Psi\rangle = \sum_{j=1}^2 c_j
|\chi_j\rangle\otimes\int \mathrm{d}\omega
\mathrm{A}(\omega)|\omega\rangle\end{equation} where $
|\chi_j\rangle, j\in\{1,2\},$ are orthonormal polarization basis
states with complex amplitudes $c_j$, and $ \mathrm{A}(\omega) $
is the complex amplitude corresponding to the frequency $\omega$,
normalized so that
\begin{equation}\label{norm1} \int\mathrm{d}\omega|\mathrm{A}(\omega)|^2=1
.\end{equation} For simplicity, we assume that the frequency
spectrum is independent of polarization, so that we need not index
$\mathrm{A}(\omega)$ by polarization mode.  In physical terms,
this means that polarization and frequency are not entangled.

Since $|\Psi\rangle$ represents a pure state, the density operator
can be written as $\rho=|\Psi\rangle\langle \Psi|$ (see
\cite{sakurai}, $\S$3.4), so that we have the initial state
\begin{eqnarray}\rho_\omega (x=0) &=& \sum_{i,j=1}^2 c_i c^{\ast}_j
|\chi_i\rangle\langle\chi_j|\nonumber\\&&
\otimes\iint\mathrm{d}\omega_1\mathrm{d}\omega_2
\mathrm{A}(\omega_1)\mathrm{A}^{\ast}(\omega_2)|\omega_1\rangle\langle\omega_2|.
\label{rho0}\end{eqnarray} The subscript indicates that
$\rho_\omega$ includes frequency degrees of freedom.  The usual
$2\times2$ density matrix representing the polarization state of
the photon is given by $\rho_{ij}=c_ic^\ast_j$.

\par We will now seek the dependence of the density operator
$\rho(x)$ on some spatially extended ``environment,'' by
introducing the operator $\mathbf{U}(x)$ which we require to be
both linear and unitary. Furthermore, we demand that
$\mathbf{U}(x)$ have eigenkets $|\chi_j\rangle$ with
(frequency-dependent) eigenvalues $U_j(\omega,x)$ where, as
before, $j\in\{1,2\}$ indexes the polarization mode. Since
$\mathbf{U}(x)$ exhibits frequency- and polarization-dependent
eigenvalues, it introduces a coupling between qubit states
(polarization modes) and non-qubit states (frequency modes).  As
shown in Appendix \ref{eigenvalues}, $\mathbf{U}(x)$ introduces
pure``phase errors" on polarization modes $|\chi_j\rangle$, so
that
\begin{equation}
\mathbf{U}(x)|\chi_j\rangle =
e^{i\varphi_j(\omega,x)}|\chi_j\rangle\label{phi}
\end{equation}
where the phase factor $\varphi_j(\omega,x)$ is a real-valued
function of $\omega$ and $x$.
\par Now we can calculate the spatial dependence of the density
operator under the influence of $\mathbf{U}(x)$ and following a
frequency-insensitive measurement, effected by a partial trace
over frequency degrees of freedom:
\begin{subequations}
\begin{eqnarray}\rho (x) &=& \int\mathrm{d}\omega
\langle\omega|\mathbf{U}(x)\rho
\mathrm{(0)}\mathbf{U^\dag}(x)\label{rho1noqc1}|\omega\rangle
\\ &=& \sum_{i,j=1}^2 c_i c^{\ast}_j
|\chi_i\rangle\langle\chi_j|\nonumber\\ &&\times
\int\mathrm{d}\omega |\mathrm{A}(\omega)|^2 e^{i [\varphi_i
(\omega,x)-\varphi_j(\omega,x)]}. \label{rho1noqc}
\end{eqnarray}
\end{subequations}

\par Since this process realizes the entangling of qubit states with
environmental states (frequency modes which are not involved in
information representation or manipulation), we expect to observe
decoherence effects in the off-diagonal elements of the qubit
state (polarization) density matrix \cite{zanardi}. In order to
investigate the coherence properties of a photon under the action
of such an operator, we observe that in a completely mixed (or
``decohered") state, the density matrix is simply a multiple of
the identity matrix. The diagonal elements of $\rho(x)$ are
indexed by $i=j$.  For these values of $i$ and $j$, the argument
of the exponential in Eq. \ref{rho1noqc} vanishes and by the
normalization condition (Eq. \ref{norm1}), we have
$\rho_{ii}(0)=\rho_{ii}(x)$.  In other words, the diagonal
elements of $\rho (x)$ are unaffected by $\mathbf{U}(x)$.
\par The off-diagonal elements of the density matrix are indexed by
the values $i=1,j=2$ and $i=2,j=1$.  To compute these explicitly,
we must define the functions $\varphi_j(\omega,x)$. In the case of
a single photon in a birefringent, linearly dispersive crystal of
thickness $x$ (e.g., a quartz crystal),
\begin{equation}\label{quartz_phi}
\varphi_j(\omega,x)=\frac{n_j x}{c} \omega
\end{equation}
where $n_j$ is the index of refraction corresponding to
polarization $j$.\footnote{Here, we neglect dispersion, the
variation of $n$ with $\omega$.  By assuming $|\chi_j\rangle$ to
be an eigenstate of $\mathcal{U}$ (Eq. \ref{phi}), we require the
optic axis of the birefringent element to be aligned with one of
the polarizations $|\chi_j\rangle$. In the case where the
$|\chi_j\rangle$ represent circular (elliptical) polarizations,
this means that the environment is optically active as opposed to
(as well as) birefringent.} Making the substitution $\tau =
\frac{(n_2-n_1)x}{c}$ and writing
$f(\omega)=|\mathrm{A}(\omega)|^2$, the off-diagonal elements of
the density matrix are given by
\begin{subequations}
\begin{eqnarray}
\rho_{12}(x)&=&\rho_{12}(0) \int\mathrm{d}\omega \mathit{f(\omega)
e^{-i \omega\tau}}\nonumber
\\&=&\rho_{12}(0) \mathcal{F^\ast(\tau)} \label{rho12}
\\ \rho_{21}(x)&=&\rho_{21}(0)\int_{-\infty}^{\infty}\mathrm{d}
\omega \mathit{f(\omega) e^{i \omega\tau}}\nonumber
\\ &=&\rho_{21}(0)\mathcal{F(\tau)} \label{rho21}
\end{eqnarray}
\end{subequations}
where $\mathcal{F(\tau)}$ is the Fourier transform of $f(\omega)$
(up to a constant, depending on convention).

\par  Evidently, $\rho^\dag(0)=\rho(0)$ implies that $\rho^\dag(x)=\rho(x)$,
so that Eqs. \ref{rho12} and \ref{rho21} preserve the required
hermiticity of $\rho$. In terms of $\Delta n=n_2-n_1$ and
$\mathcal{F}(\tau)$, the density matrix at some later position $x$
becomes a simple modification of the density matrix at $x=0$:
\begin{equation}
\rho(x)=
\begin{pmatrix}\rho_{11}(0)&\rho_{12}(0)\mathcal{F}^\ast(\frac{x\mathrm{\Delta}
\mathit{n}}{c}) \cr \rho_{21}(0)\mathcal{F}(\frac{x\mathrm{\Delta}
\mathit{n}}{c})&\rho_{22}(0)\label{rhonoqc}
\end{pmatrix}.
\end{equation}
\par Eq. \ref{rhonoqc} allows us to calculate the density
matrix representing the polarization of an optical qubit subjected
to a non-dissipative frequency-polarization coupling, given the
initial state density matrix, $\rho(0)$, (written in the basis of
eigenmodes of the coupling) and the functional form of the
frequency-amplitude function
$\mathrm{A}(\omega)$.\footnote{Actually, we need only know the
complex square of the frequency-amplitude function,
$|\mathrm{A}(\omega)|^2$.} In general, $|\mathrm{A}(\omega)|^2$ is
peaked at some central value $\omega_o$ with a finite width
$\delta\omega$.  Under these conditions, the Fourier transform
$\mathcal{F}(\tau)$ (and the off-diagonal elements of $\rho$) will
fall significantly on the time scale
$\tau_c\sim\frac{1}{\delta\omega}$. In other words, there is no
coherent phase relationship between polarization basis states
$|\chi_1\rangle$ and $|\chi_2\rangle$ after the photon has
traveled a distance $x>\frac{c}{\delta\omega}$ through the
environment $\mathbf{U}(x)$.

\par To summarize, we have developed a prescription for calculating
the polarization state density matrix representing a photon in a
birefringent environment.  From this expression, we see that a
photon with definite polarization phase information before
entering such an environment will, in general, lose phase
information between polarization basis states after traveling a
suitably long distance.  The physical mechanism which destroys
such phase information is the entanglement between polarization
modes and frequency modes, which is an instance of the
entanglement between qubit states and environmental degrees of
freedom which are not useful for information manipulation. Because
we trace over frequency (environmental) degrees of freedom in this
case, the phase coherence between the qubit basis states is
destroyed.

\subsection{``Bang-bang" quantum control of
decoherence}\label{bang-bang}

\par In the previous section, we showed that a qubit consisting of
photon polarization modes $|\chi_j\rangle$ may lose its capacity
for information representation under the action of a unitary
operator, $\mathbf{U}(x)$, which entangles frequency and
polarization. In other words, decoherence occurs even for
non-dissipative ``phase errors'' in which the qubit basis states
are eigenstates of the environmental coupling, and there is no
chance for a bit flip error \cite{unruh}.\footnote{Note however,
that phase errors in the basis $\{|\chi_j\rangle\}$ may appear as
bit-flip errors in another basis, e.g. $\{\frac{(|\chi_1\rangle
\pm |\chi_2\rangle)}{\sqrt{2}}\}$.}

\par In \cite{bang-bang}, the authors describe this result in the
general case of qubit states coupled to environmental degrees of
freedom and introduce a scheme for reducing such decoherence
effects via rapid, periodic exchanging of the eigenstates of the
environmental coupling. In our case, this corresponds to an
interchange of the polarization basis states $|\chi_1\rangle$ and
$|\chi_2\rangle$, which can be accomplished by an appropriate
reflection or rotation operation. These exchanges are rapid in the
sense that the period of flipping must be short compared to the
decoherence time $\tau_c \sim \frac{1}{\delta\omega}$ and the time
scales of any other relevant decoherence mechanisms (e.g.
scattering, dissipation). So called ``bang-bang'' control reduces
the degree of decoherence by averaging out the time-dependent
coupling between qubit states and environmental states.

\par In this section, the previous results are
expanded to include an implementation of ``bang-bang'' quantum
control by periodic exchanging of the environmental eigenstates
$|\chi_j\rangle$. The mathematical formulation is considerably
simplified if we move to the discrete regime in which the
``environment'' acts in a stepwise fashion, and the exchange of
eigenstates also occurs in discrete steps.\footnote{Developing the
general continuous case in analogy with Sect. \ref{review} by
including a rotation operator acting simultaneously with
$\mathbf{U}(x)$ is considerably complicated, since this rotation
operator will not, in general, commute with $\mathbf{U}\mathit
(x)$ for different $x$ values.} To this end, we define ${\bf R}$
to be a reflection (or rotation) operator which exchanges the
eigenstates of the frequency-polarization coupling:
\begin{subequations}
\begin{eqnarray}
\mathbf{R}\mathit|\chi_1\rangle &=& |\chi_2\rangle\label{R1}\\
\mathbf{R}\mathit|\chi_2\rangle &=&\pm|\chi_1\rangle.\label{R2}
\end{eqnarray}
\end{subequations}
Here, the plus (minus) sign refers to a reflection (rotation). The
choice of sign will not affect the results of a measurement in the
experimental case, and for concreteness, we take the sign to be
positive.

\par We define $\mathcal{U}$, a ``step-wise" operator analogous
to $\mathbf{U}(x)$ in Sect. \ref{review} and also define
$\mathcal{L}$ to be the thickness of the birefringent element (in
our experiments, usually a quartz crystal); as before, $n_j$ is
the refractive index corresponding to polarization state $j$, and
$\Delta n=n_2-n_1$.  This gives a relative phase shift of $\varphi
(\omega) \equiv \varphi_2 (\omega)-\varphi_1(\omega) = \left(
\frac{\mathcal{L}\Delta n}{c} \right)\omega$, so that we
have\footnote{Here, we omit the phase shift of $\varphi_1(\omega)=
\left(\frac{\mathcal{L}n_1}{c}\right)\omega$, which is common to
Eqs. \ref{scriptU1} and \ref{scriptU2}. Such a global phase shift
is never observable and cannot affect the coherence properties of
the polarization states.}
\begin{subequations}
\begin{eqnarray}
\mathcal{U}|\chi_1\rangle&=&|\chi_1\rangle\label{scriptU1}\\
\mathcal{U}|\chi_2\rangle&=&e^{i\varphi(\omega)}|\chi_2\rangle.
\label{scriptU2}
\end{eqnarray}
\end{subequations}
\par Eqs. \ref{R1}-\ref{R2} and \ref{scriptU1}-\ref{scriptU2} have the important
consequence that
\begin{subequations}
\begin{eqnarray}
\left(\mathbf{R}\mathcal{U}\right)^2&=&e^{i \varphi
(\omega)}\label{identity1}\\
\left(\mathcal{U^\dag\mathbf{R^\dag}}\right)^2&=&e^{-i \varphi
(\omega)}\label{identity2}
\end{eqnarray}
\end{subequations}
where it is understood that such operator identities are only
meaningful when the operators are applied to kets or bras (see
Appendix \ref{proof1} for a proof of these identities).

\par In this discrete regime, we consider ``rapid'' exchanging of
eigenstates to correspond to the alternating action of the
operators $\mathcal{U}$ and $\mathbf{R}$.  Regarding $\mathcal{U}$
and $\mathbf{R}$ as optical elements (for example, a quartz
crystal and a half-wave plate), we consider a cavity scheme in
which a single photon passes through this two-element unit $N$
times.
\par Symbolically, after $N$ passes through the system, we have
(compare Eq. \ref{rho1noqc1}),
\begin{equation}
\rho (N)=\int
\mathrm{d}\omega\langle\omega|\left(\mathbf{R}\mathcal{U}\right)^N
\rho_\omega (0)\left(
\mathcal{U^\dag}\mathbf{R^\dag}\right)^N|\omega\rangle\label{rho1qc1}.
\end{equation}

Application of identities \ref{identity1} and \ref{identity2}
leads to a simplification, depending on whether $N$ is odd
($N=2m+1$) or even ($N=2m$).
\begin{subequations}
\begin{eqnarray} \rho (2m)&=&\int \mathrm{d}\omega\langle\omega|\left(\mathbf{R}
\mathcal{U}\right)^{2m} \rho_\omega (0)\left(
\mathcal{U^\dag}\mathbf{R^\dag}\right)^{2m}|\omega\rangle\nonumber\\
&=&\int\mathrm{d}\omega\langle\omega|e^{2m i \varphi (\omega)}
\rho_\omega (0)e^{-2m i \varphi (\omega)}|\omega\rangle\nonumber\\
&=& \int\mathrm{d}\omega\langle\omega|\rho_\omega
(0)|\omega\rangle\nonumber\\ &=&\rho(N=0)\\ \rho (2m+1)&=& \int
\mathrm{d}\omega\langle\omega|\left(\mathbf{R}
\mathcal{U}\right)^{2m+1}\rho_\omega (0)\left(
\mathcal{U^\dag}\mathbf{R^\dag}\right)^{2m+1}|\omega\rangle
\nonumber\\ &=&\int \mathrm{d}\omega\langle\omega|e^{2m i \varphi
(\omega)} \mathbf{R}\mathcal{U}\rho_\omega (0)
\mathcal{U^\dag}\mathbf{R^\dag}e^{-2m i \varphi
(\omega)}|\omega\rangle\nonumber\\&=&\int\mathrm{d}\omega\langle\omega|
\mathbf{R}\mathcal{U}\rho_\omega (0)
\mathcal{U^\dag}\mathbf{R^\dag} |\omega\rangle\nonumber\\ &=&
\rho(N=1).
\end{eqnarray}
\end{subequations}
Note that $\rho(N=1)$ is related to the continuous case of Sect.
\ref{review} by
\begin{equation}
\rho(N=1) = \int{\rm d}\omega\langle\omega|{\bf R}\rho(x={\cal L})
{\bf R}^\dag |\omega\rangle.
\end{equation}
These relations lead to a simple analog of Eq. \ref{rho1noqc}
(where we use $\rho_\omega(N=0)\leftrightarrow\rho_\omega(x=0)$,
given by Eq. \ref{rho0}, as the input state in both cases):
\begin{subequations}
\begin{eqnarray}
\rho(N=0)&=&\sum_{i,j=1}^2c_ic_j^\ast|\chi_i\rangle\langle\chi_j|
\\ \rho(N=1)&=&\sum_{i,j=1}^2c_ic_j^\ast\int\mathrm{d}\omega|A
(\omega)|^2\left(\mathbf{R}\mathcal{U}\right)|\chi_i\rangle\langle\chi_j|
\left(\mathcal{U^\dag}\mathbf{R^\dag}\right).\nonumber\\
\end{eqnarray}
\end{subequations}

\par Finally, writing out $\rho(N)$ in matrix form, substituting
$\mathcal{F}$ for the Fourier transform of
$f(\omega)=|A(\omega)|^2$, and denoting by
$\tau_0=\frac{\mathcal{L}\Delta n}{c}$ the characteristic time
parameter, we have
\begin{subequations}
\begin{eqnarray}
\rho_{\scriptscriptstyle{QC}}(2m)&=&\begin{pmatrix}\rho_{11}(0)&\rho_{12}(0)
\cr \rho_{12}^\ast(0)&\rho_{22}(0)\end{pmatrix}\label{rhoqc1}\\
\rho_{\scriptscriptstyle{QC}}(2m+1)&=&\begin{pmatrix}\rho_{11}(0)&\rho_{12}^\ast(0)
\mathcal{F} \left(\tau_0\right) \cr \rho_{12}(0)\mathcal{F}^\ast
\left(\tau_0\right)&\rho_{22}(0)\end{pmatrix}\label{rhoqc2}.\nonumber\\
\end{eqnarray}
\end{subequations}

\par The subscript ${\scriptscriptstyle{QC}}$ (for ``Quantum Control'')
indicates the case in which $\mathbf{R}$ is included. Note that
$\rho_{\scriptscriptstyle{QC}}$ is dependent only on the parity of
$N$, so that under the influence of the quantum control procedure,
any number of cycles through this particular birefringent
environment causes {\em at most} a decoherence equal to that of
the first pass.  In fact, Eq. \ref{rhoqc1} shows that an even
number of cycles causes no decoherence whatsoever.

\par By making the identification $x = N\mathcal{L}$ (the distance
traveled through the crystal after N passes) in Eq. \ref{rhonoqc},
we can directly compare the results of Sect. \ref{review} with
those of Sect. \ref{bang-bang}:
\begin{equation}\label{rhonoqcmatrix}
\rho(N)=
\begin{pmatrix}\rho_{11}(0)&\rho_{12}(0)\mathcal{F}^\ast\left(N\tau_0\right) \cr
\rho_{12}^\ast(0)\mathcal{F}\left(N\tau_0\right)&\rho_{22}(0)
\end{pmatrix}.
\end{equation}

\par Letting $\tau_c$ be the coherence time such that
$\mathcal{F}(\tau)$ is significantly reduced for $\tau>\tau_{c}$,
then $\rho(N)$ represents a partially mixed state for
$N>\frac{\tau_c}{\tau_0}$.\footnote{Here, ``mixed" refers to the
fact that the off-diagonal elements of $\rho$ are small so that
$Tr(\rho^2)<1$. If the initial state, $\rho(0)$, has diagonal
elements of $1/2$, then $\rho$ represents a completely random
ensemble for $N\gg\frac{\tau_c}{\tau_0}$.} On the other hand,
$\rho_{\scriptscriptstyle{QC}}(N)$ never loses all phase
information in the off-diagonal elements since $N=2m$ recovers the
initial state for all $m$.  In this sense, $\mathbf{R}$ acts as a
quantum control element, suppressing decoherence and maintaining
state purity indefinitely.

\par We may ask what happens if the strength of the environmental
coupling varies with $x$, so that $\mathcal{U}$ varies on a time
scale comparable to the period of the exchange operator,
$\mathbf{R}$. For simplicity, we model this ``slow-flipping'' case
by introducing two distinct operators which differ only in their
eigenvalues (not their eigenstates). In more concrete terms, we
consider two coupling operators, $\mathcal{U}_1$ and
$\mathcal{U}_2$, characterized by phase errors $\varphi_1(\omega)$
and $\varphi_2(\omega)$ so that, for $j = 1,2$:
\begin{subequations}
\begin{eqnarray}
\mathcal{U}_j|\chi_1\rangle&=&|\chi_1\rangle\label{scriptU11}\\
\mathcal{U}_j|\chi_2\rangle&=&e^{i\varphi_j(\omega)}|\chi_2\rangle.\label{scriptU22}
\end{eqnarray}
\end{subequations}
\par The exchange operator ${\bf R}$ acts in alternation between each of
these two operators, so that the strength of the coupling changes
faster than the eigenstates of the coupling are exchanged.  After
N passes through the system, we have
\begin{subequations}
\begin{eqnarray}
\rho(N) & = & \int\mathrm{d}\omega\langle\omega|(\mathcal{U}_2
\mathcal{U}_1)^N\rho_\omega(0)(\mathcal{U}_1^\dag
\mathcal{U}_2^\dag )^N|\omega\rangle\\
\rho_{\scriptscriptstyle{QC}}(N) & = &
\int\mathrm{d}\omega\langle\omega|(\mathbf{R}\mathcal{U}_2
\mathbf{R}\mathcal{U}_1)^N\rho_\omega(0)(\mathcal{U}_1^\dag
\mathbf{R}^\dag\mathcal{U}_2^\dag
\mathbf{R}^\dag)^N|\omega\rangle.\nonumber\\
\end{eqnarray}
\end{subequations}

Intuitively, we expect to see a partial reduction in the degree of
decoherence when the quantum control procedure is implemented by
introducing the exchange operator.

\par To see these results quantitatively, we make the following substitution
which reduces this problem (mathematically) to the previously
solved one (see Appendix \ref{proof2} for a derivation of these
relations):
\begin{subequations}
\begin{eqnarray}
\mathcal{U}_2\mathcal{U}_1&\equiv&\overline{\mathcal{U}}\\
\mathbf{R}\mathcal{U}_2\mathbf{R}\mathcal{U}_1&\equiv&
\overline{\mathcal{U}}_{\scriptscriptstyle{QC}}
\end{eqnarray}
\end{subequations}
so that,
\begin{subequations}
\begin{eqnarray}
\overline{\mathcal{U}}|\chi_1\rangle & =
&|\chi_1\rangle\label{relation1}\\
\overline{\mathcal{U}}|\chi_2\rangle & = &
e^{i\left(\varphi_1(\omega)+\varphi_2(\omega)\right)}|\chi_2\rangle\label{relation2}
\end{eqnarray}\end{subequations}\begin{subequations}\begin{eqnarray}
\overline{\mathcal{U}}_{\scriptscriptstyle{QC}}|\chi_1\rangle &=&
|\chi_1\rangle\label{relations3}\\
\overline{\mathcal{U}}_{\scriptscriptstyle{QC}}|\chi_2\rangle &=&
e^{i\left(\varphi_1(\omega)-\varphi_2(\omega)\right)}|\chi_2\rangle.\label{relation4}
\end{eqnarray}
\end{subequations}
\par With these relations, each operator is mathematically analogous to
the operator $\mathcal{U}$ defined by Eq. \ref{phi}, with
$\varphi(\omega)$ redefined as appropriate given relations
\ref{relation1}-\ref{relation4} and the details of the particular
environment.  In the case of birefringent crystals, the elements
corresponding to $\mathcal{U}_1$ and $\mathcal{U}_2$ differ only
in their thickness (not in their orientation with respect to the
polarization basis states $|\chi_j\rangle$). Defining
$\mathcal{L}_j$ to be the thickness of a crystal represented by
$\mathcal{U}_j$, we have
\begin{equation}\label{phi_eff}
\varphi_j(\omega) = \left(\frac{\mathcal{L}_j\Delta
n}{c}\right)\omega.
\end{equation}
\par We see that before introducing the exchange operator, the
birefringent environment acts effectively as a single birefringent
crystal of thickness $\mathcal{L}_1+\mathcal{L}_2$ (Eq.
\ref{relation2}). After implementing ``bang-bang'' control by
introducing the exchange operator $\mathbf{R}$, the environment
looks effectively like a single crystal of thickness
$\mathcal{L}_1-\mathcal{L}_2$ (Eq. \ref{relation4}). In the
absence of quantum control, the photon is subjected to the {\em
sum} of the individual phase errors; in the presence of a quantum
control element, the photon is subjected to the {\em difference}
of the individual phase errors.  In this case, where the strength
of the environmental coupling changes on roughly the same time
scale as the exchanging of eigenstates, the quantum control
procedure reduces but does not eliminate decoherence effects.

\subsection{Experimental Demonstration of ``Bang-Bang'' Control}\label{experiment}

\par It will be convenient in our experimental analysis
to measure the \textit{visibility}, traditionally defined as
\begin{equation}
{\cal V} = \frac{N_{max}-N_{min}}{N_{max}+N_{min}}\label{V_old}
\end{equation}
where $N_{max}$ ($N_{min}$) denotes the maximum (minimum) number
of photons passing through orthogonal settings of a polarizer.
Denoting these polarization analyzer settings by $|\lambda\rangle$
and $|\overline{\lambda}\rangle$, we define visibility at the
single quantum level by calculating probabilities from the density
operator, $\rho(x)$\footnote{The notation
$\{|\lambda\rangle,|\overline{\lambda}\rangle\}$, is intended to
suggest that these polarization modes may, in general, represent
any elliptical orthonormal basis states.}:
\begin{equation}\label{V_lambda}
{\cal V}_\lambda(x) =
{|\langle\lambda|\rho(x)|\lambda\rangle-\langle\overline{\lambda}|\rho
\mathit{(x)}|\overline{\lambda}\rangle|}.
\end{equation}
Note that we no longer include the denominator corresponding to
the total number of photons in Eq. \ref{V_old}, since the
denominator here is the trace of $\rho$ in the
$\{|\lambda\rangle,|\overline{\lambda}\rangle\}$ basis, which is
unity. $\mathcal{V}_\lambda(x)$ quantifies the similarity between
the state represented by $\rho (x)$ and the state
$|\lambda\rangle$ in the following sense: if $\rho (x)$ represents
an unpolarized state, it is a multiple of the identity matrix, and
$\mathcal{V}_\lambda(x)=0$; at the opposite extreme, if the photon
is in a pure state with polarization $\lambda$, $\rho (x) =
|\lambda\rangle\langle\lambda|$, and ${\cal V}_\lambda(x) = 1$. If
we choose to measure with respect to the state $|\lambda\rangle =
\frac{1}{\sqrt{2}}\left(|\chi_1\rangle+|\chi_2\rangle\right)$, we
find that (See Appendix \ref{vis_proof} for a proof)
\begin{equation}\label{visibility}
\mathcal{V}(x)=2Re(\langle\chi_1|\rho(x)|\chi_2\rangle)
\end{equation}
where we have dropped the subscript $\lambda$ for this choice of
measurement, and $Re$ indicates the real part. In this form, it is
clear that visibility ${\cal V}(x)$ depends on the off-diagonal
elements of $\rho(x)$, and so corresponds with our previous notion
of coherence. If we write $\rho$ in the $H/V$ basis, measuring the
visibility with respect to
$|\lambda\rangle=\frac{1}{\sqrt{2}}(|H\rangle+|V\rangle)=|45^\circ\rangle$
and
$|\overline{\lambda}\rangle=\frac{1}{\sqrt{2}}(|H\rangle-|V\rangle)=|-45^\circ\rangle$
gives us information about the off-diagonal elements of $\rho$.

\par
\begin{figure}[t]
\begin{center}\scalebox{.35}[.35]{\includegraphics{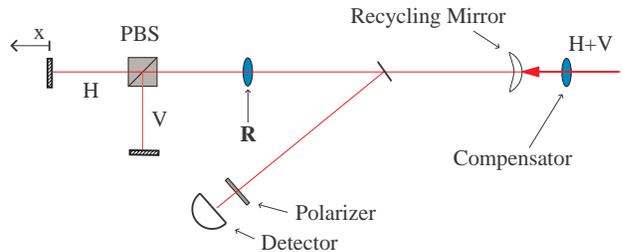}}\end{center}
    \caption{\label{michelson}Schematic diagram of experimental apparatus for producing
    controllable decoherence in the H/V basis.  PBS indicates a
    Polarizing Beam Splitter; $\mathbf{R}$ indicates the (removable)
    exchange element, a quarter-wave plate in this case.  The input
    state is $|45^\circ\rangle \left(\equiv\frac{|H\rangle+|V\rangle}
    {\sqrt{2}}\right)$.}
\end{figure}

 To see the results of the previous sections experimentally,
we seek a systematic and controllable implementation of the
operators $\mathcal{U}$ and $\mathbf{R}$ in the cavity scheme of
Sect. \ref{bang-bang}.  For the first case, in which there is a
single operation which produces a phase error, we use a Michelson
polarization interferometer as a tunable decoherence mechanism in
the $H/V$ basis. A polarizing beam-splitter (PBS) separates the
path of $V$- and $H$-polarized light.  The unbalanced arms of the
interferometer produce a controllable, $\omega$-dependent phase
shift between $|H\rangle$ and $|V\rangle$ (see Fig.
\ref{michelson}).
\par A photon in either path is reflected back to the PBS and retraces
its path to the input. The difference in path lengths between arms
of the interferometer will be denoted by $x$.  The unbalanced
interferometer is the optical realization of the operator
$\mathbf{U}$ of Sect. \ref{review} and $\mathcal{U}$ of Sect.
\ref{bang-bang}.

\par An $R=0.97$ reflector at the input recycles the light so
 that a photon passes through the unbalanced arms of the interferometer more
than once, in general.\footnote{We use $R$ to denote the {\em
probability} of reflection at a surface.} A partially reflective
out-coupling mirror directs a photon out of the interferometer
with a probability of $R=0.04$ at each pass.\footnote{For normal
incidence, $R=0.04$ at each air-glass interface.  Here, one
surface of the out-coupling mirror is broad-band anti-reflection
(BBAR) coated so that we have effectively only a single air-glass
interface.}

\par We send in light from a greatly attenuated diode laser at $\lambda = 670$
nm, pulsed at $100$ kHz. In our experiment, we observe photons
passing through the interferometer $\leq 10$ times.  The total
distance traveled by a photon passing through our interferometer
$10$ times is $\sim5$ m, so that any photon spends at most
$\sim10^{-8}$ s in the system. EG\&G SPCM-AQ silicon avalanche
photodiodes (dark count $<400\ {\text s}^{-1}$, efficiency $\sim60
\%$) are used as photon counters. Using electronic timing
techniques, we record photon counts only in a coincidence window
of 50 ns between the pulse generator and detection event. These
timing techniques, together with spectral filters at the
detectors, reduce background coincidence counts to $<1$ s$^{-1}$.
We graphically display photon counts vs. arrival time so that
photons leaving the system after N cycles are recorded in the Nth
peak. In a typical experiment, our photon counters register no
more than $5000$ coincidence counts per second, and we conclude
that $\sim10,000$ photons per second leave the system ($\sim 0.1$
per pulse).\footnote{The probability that there are two photons in
the system simultaneously is given by the Poisson distribution,
$P(2)=\frac{e^{-0.1}(0.1)^2}{2!}\sim0.005$. These photon numbers
are comparable to those used in current demonstrations of
free-space quantum cryptographic key distributions \cite{crypto}.}
At the output, we count for a few seconds (less than a minute in
most cases). Although we cannot determine the unknown polarization
state of a single photon, we can accurately find the statistical
distribution of the ensemble over the observation time, from which
useful information about the density matrix can be inferred.

\par  By placing a $\lambda/4$ waveplate with optic axis at $45^\circ$
at the recycling reflector, we realize an optical ``flipping"
operator.  The $\lambda/4$-reflector-$\lambda/4$ combination takes
$|H\rangle$ to $|V\rangle$ and {\em vice versa}. This is the
optical realization of the exchange operator, $\mathbf{R}$ in
Sect. \ref{bang-bang}.

\par By recording photon counts in coincidence with the diode
pulse generator and using a polarizer at the detector, we can
measure the visibility at
$|45^\circ\rangle=\frac{1}{\sqrt{2}}(|H\rangle+|V\rangle)$. The
frequency spectrum of laser light is restricted using an
interference filter with bandwidth $\delta\lambda =10$ nm and
central wavelength $\lambda_o = 670$ nm.\footnote{Since the diode
laser is weakly driven just above its lasing threshold, the
intrinsic frequency spectrum of the photons is wide enough that
these filters determine the overall frequency distribution.} For
our calculations, we use a Gaussian amplitude function with
frequency spread $\delta\omega$ corresponding to $\delta\lambda
=10$ nm, central frequency $\omega_0$ corresponding to $\lambda_o
=670$ nm, and normalization factor $A_o$:\footnote{In practice, we
find that the frequency spectrum is not as smooth as a Gaussian.
From the visibility curves, we infer that the frequency spectrum
has a substantial monochromatic (compared to $\delta\omega$)
component at $\omega_o$.\label{monofootnote}}
\begin{equation}
A(\omega) = A_o
\exp\left[-\left(\frac{\omega_{\scriptscriptstyle{0}}
-\omega}{\delta\omega/\sqrt{2}}\right)^2\right].
\end{equation}
The corresponding coherence length of the photons is given by
$L_c\sim\frac{\lambda_o^2}{\delta\lambda}=45\mu$m, so we expect
substantial decoherence for a path difference of $\sim4.5\ \mu$m
at cycle $N=10$.
    \begin{figure}
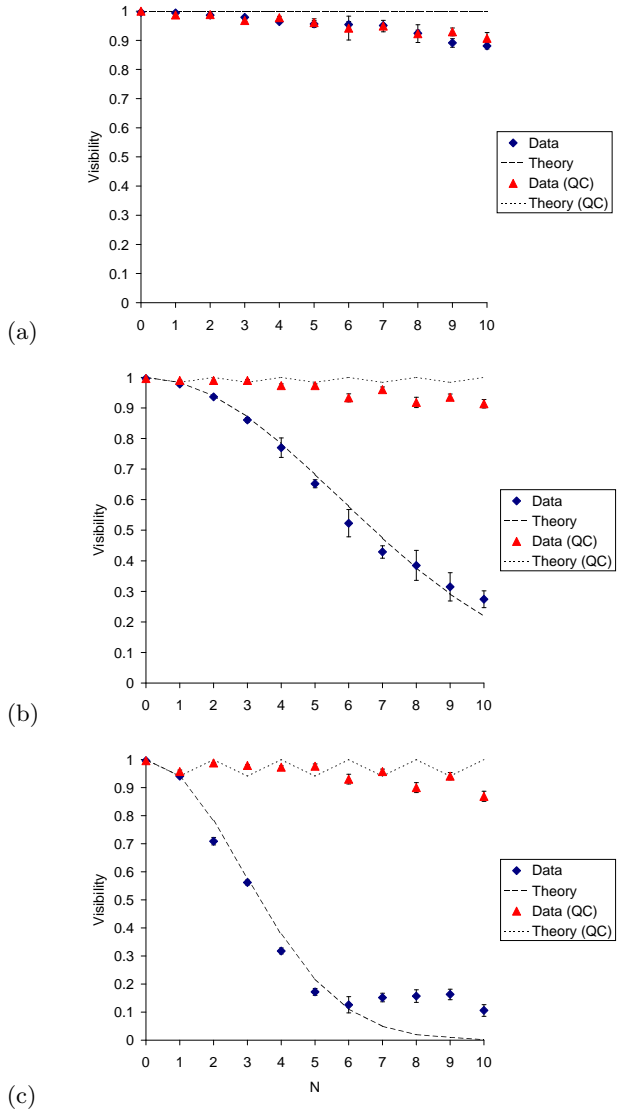

    \begin{tabular}{lc}
(a) & \scalebox{0.3}[0.3]{\includegraphics*{0.00microns.prn}}
\\ (b) & \scalebox{0.3}[0.3]{\includegraphics*{2.49microns.prn}}
\\ (c) & \scalebox{0.3}[0.3]{\includegraphics*{4.99microns.prn}}
    \end{tabular}
    \caption{Experimental and theoretical curves showing
    $\mathcal{V}$ vs. $N$ for the case where (a) $x=0.0\ \mu m$, (b) $x=2.49\ \mu m$,
    and (c) $x=4.99\ \mu m$, for the setup in Fig. \ref{michelson}. ``QC''
    indicates the case in which the quantum control
    procedure was implemented. In (a) ($x=0$), the theoretical curves
    for both cases are constant at unity, since we expect no decoherence in this
    case.  In (c), the observed visibility curve (without quantum control)
    falls to $\mathcal{V} \sim 0.15$ on the length scale of our
    observations. We suspect that $\mathcal{V}$ falls to 0 on the
    length scale determined by the monochromatic component of the
    frequency spectrum mentioned in footnote \ref{monofootnote}.
    The experimentally observed values in the quantum control case
    alternate in the opposite sense to the theoretical values.
    This discrepancy is not yet understood.}
    \label{0.00microns}
    \end{figure}
\par With this setup, we can measure $\mathcal{V}$ vs. N as in
Sect. \ref{bang-bang}.  Experimental and theoretical results are
plotted in Fig. \ref{0.00microns} for the case where the arms are
unbalanced by $x=0.00,\ 2.49\text{, and }4.99 \mu$m. The
experimental results show clearly that the introduction of the
quantum control element results in a significant reduction of
decoherence in good agreement with the theory. However, practical
difficulties with the interferometer limit the quantitative
agreement between experimental and theoretical values of
$\mathcal{V}_{\scriptscriptstyle{QC}}$, especially at large N. We
see in Fig. \ref{0.00microns}(a) that the relative path difference
is not 0 after $N=10$ passes through the interferometer due to
imperfect balancing. By determining the coherence length from the
data in Fig. \ref{0.00microns}, we infer that the actual path
length difference in the interferometer was $\sim\pm0.7\ \mu$m
(about one wavelength of incident light) displaced from the
nominal $x$-values. This discrepancy then determines the bounding
value of $\mathcal{V}(N=10)$ for all other plots. Theoretically,
of course, this value is exactly unity.
\par
\begin{figure}
    \scalebox{.35}[.35]{\includegraphics{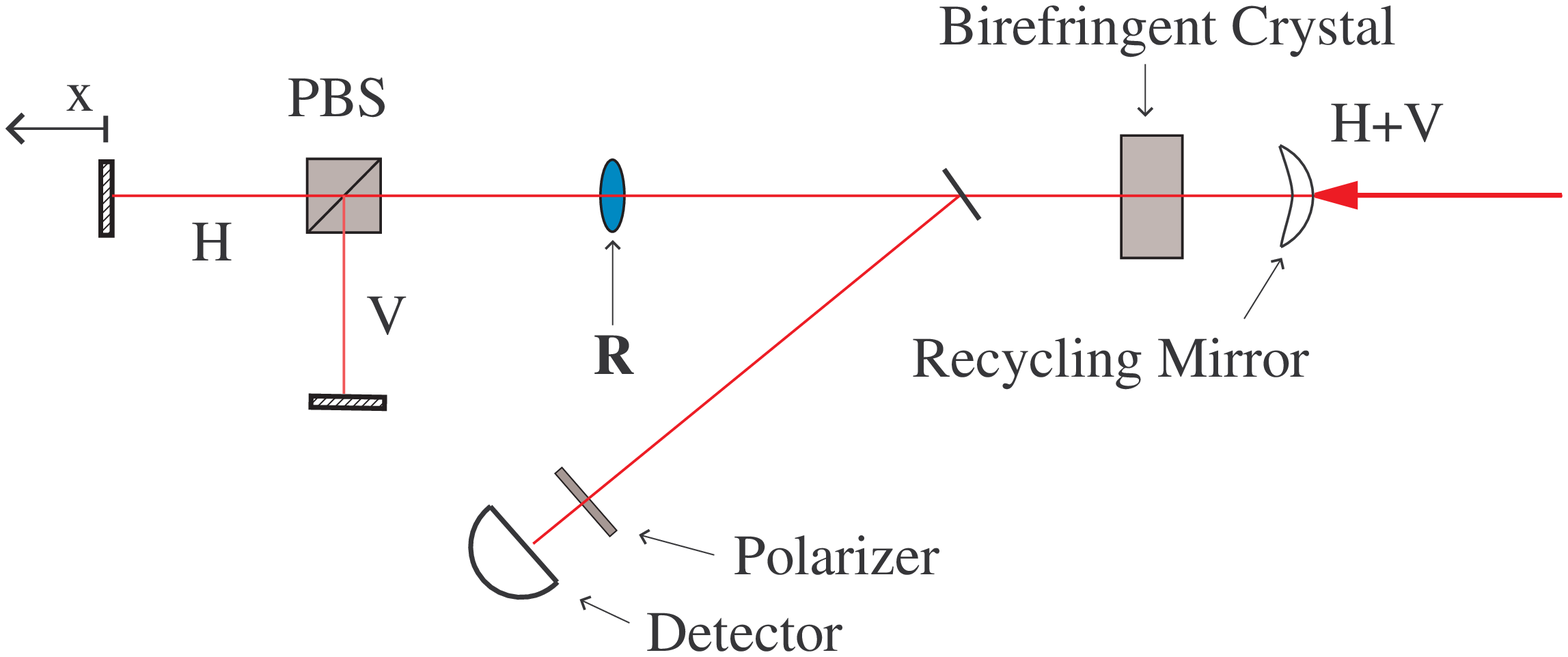}}
    \caption{\label{michelson_slow}Schematic diagram of
    experimental apparatus for the ``slow-flipping'' case. As before,
    $\mathbf{R}$ indicates the exchange element, a half-wave plate in
    this case.  The birefringent crystal, a quartz plate, introduces a
    second phase error in the same basis as the unbalanced
    interferometer arms.}
    \end{figure}

\begin{figure}
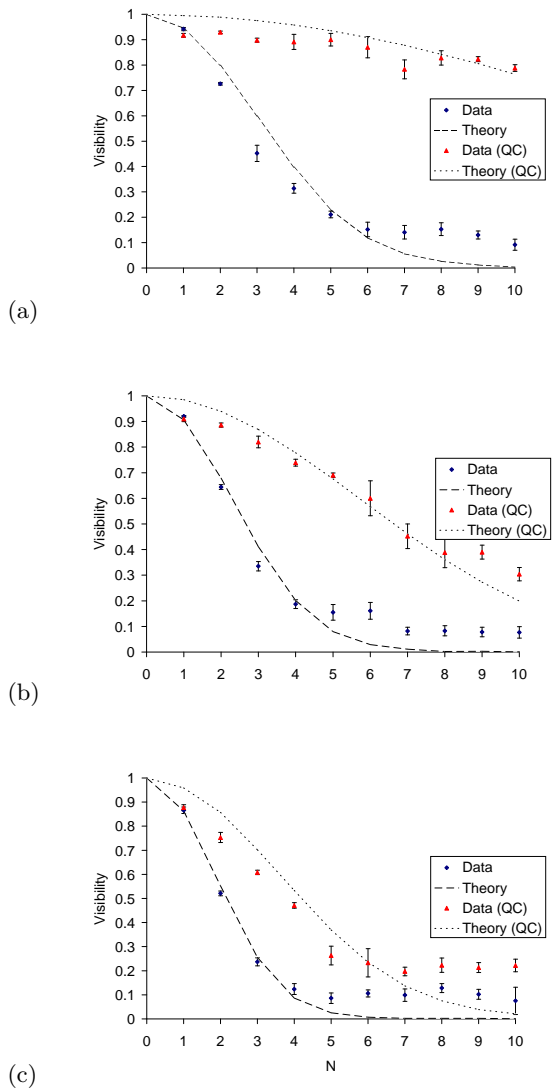

    \begin{tabular}{lc}
(a) & \scalebox{0.3}[0.3]{\includegraphics*{2.1slow.prn}}
\\ (b) & \scalebox{0.3}[0.3]{\includegraphics*{3.16slow.prn}}
\\ (c) & \scalebox{0.3}[0.3]{\includegraphics*{4.2slow.prn}}
    \end{tabular}
    \caption{Experimental and theoretical curves showing
    $\mathcal{V}$ vs. $N$ for the ``slow-flipping'' case where $x_Q=1.77\ \mu$m
    and (a) $x_1=2.1\ \mu m$, (b) $x_2=3.16\ \mu m$, and (c) $x_3=4.2\ \mu m$
    for the setup in Fig. \ref{michelson_slow}.}
    \label{slow}
    \end{figure}

 In order to model the ``slow-flipping'' case, we
introduce an additional operator, $\mathcal{U}_2$, which will
induce a second phase shift, also in the $H/V$ basis.  To
accomplish this task, we include a birefringent quartz crystal
whose optic axis is aligned with the axes of the PBS of the
original setup.  We denote the optical path difference in this
crystal by $x_Q$.\footnote{\label{1.77_footnote} Instead of a
single quartz crystal, we use a 1.046 mm crystal at $0^\circ$ and
a 0.850 mm crystal perpendicular to the first. The combination of
these two crystals gives an effective path length such that nearly
complete decoherence occurs over 10 cycles.} We now use a
$\lambda/2$-waveplate as the quantum control element which
exchanges the eigenstates, $H$ and $V$, at each pass both to the
left and to the right (see Fig. \ref{michelson_slow}). For the
theoretical values, we used the same prescription as above, but
with effective operators defined by Eqs.
\ref{relation1}-\ref{relation4}. Experimental and theoretical
results are displayed in Fig. \ref{slow}.

\par
\begin{figure}
    \scalebox{.3}[.3]{\includegraphics{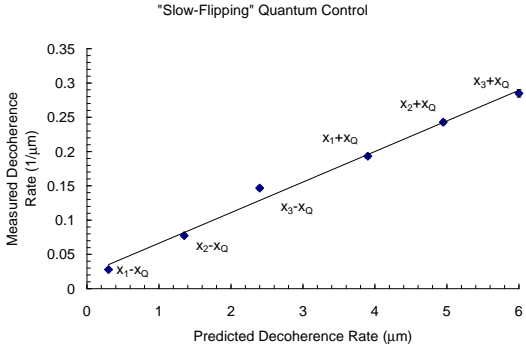}}
    \caption{\label{summary_plot}Measured (reciprocal) optical path
    difference inferred from the coherence lengths in Fig. \ref{slow}
    versus expected values according to Sect. \ref{bang-bang}.
    Error bars in both directions are within the points.
    See the text for an explanation.}\end{figure}

 In Fig. \ref{summary_plot}, we have inferred the effective rate
of decoherence (a reciprocal length, in this case) in the system
by fitting curves to each of Fig. \ref{slow} and (reciprocally)
plotted these versus the rate of decoherence for a single pass
predicted by Eqs. \ref{relation2} and \ref{relation4}. We expect
to see an effective decoherence rate given by the {\em sum} of
$x_Q$, the decoherence rate due to the quartz crystal, and $x_i$,
$i\in\{1,2,3\}$, the decoherence rate due to the interferometer,
when the quantum control procedure is not implemented;
furthermore, we expect to see a decoherence rate determined by the
{\em difference} of $x_Q$ and $x$ when the quantum control
procedure is implemented. Here, the decoherence ``rate'' of an
optical element is directly proportional to the reciprocal of the
optical path difference between e- and o-polarizations. The
linearity of this plot indicates that the quantum control element
acts as predicted in Sect. \ref{bang-bang}.

\par We are limited in the type of decoherence we can
observe by measuring ${\cal V}$. For example, a photon in the pure
state $|H\rangle$ has a visibility of $0$ with respect to the
$45^\circ$ basis. We would like a more general method of
quantifying decoherence effects. To this end, we can extend the
various methods of polarization analysis in classical optics to
our more fundamental quantum mechanical setting. With this idea in
mind, we define the {\em degree of polarization} $\mathcal{P}$,
and the Stokes parameters $\langle s_i\rangle$ in terms of $\rho
(x)$ as\footnote{These are a generalization of the Stokes
parameters of classical optics to the quantum mechanical case. In
classical optics, these data can be used to reconstruct the
coherency matrix, $\mathbf{J}$, which completely characterizes the
polarization state of the light (See \cite{born_and_wolf}, $\S
10.9$).}
\begin{equation}
\mathcal{P}\mathrm = \sqrt{\langle s_1\rangle ^2+\langle s_2
\rangle ^2+\langle s_3 \rangle ^2}\label{purity}
\end{equation}
and
\begin{subequations}
\begin{eqnarray}\label{stokes1}\langle s_1 \rangle &=&2\langle H|\rho (x)|H\rangle-1
\\ \langle s_2 \rangle &=&2\langle 45^\circ|\rho (x)|45^\circ\rangle-1 \nonumber
\\   &=&\langle H|\rho (x) |V\rangle +\langle V|\rho (x)|H\rangle
\\ \label{stokes2}\langle s_3 \rangle&=&2\langle R|\rho (x)|R\rangle-1 \nonumber
\\   &=& i (\langle H|\rho (x) |V\rangle-\langle V|\rho
(x)|H\rangle).\label{stokes3}
\end{eqnarray}
\end{subequations}
As usual, $|H\rangle$ and $|V\rangle$ denote horizontal and
vertical polarization, $|45^\circ\rangle =
\frac{1}{\sqrt{2}}(|H\rangle+ |V\rangle)$ denotes $45^\circ$
polarization, and $|R\rangle = \frac{1}{\sqrt{2}}(|H\rangle+ i
|V\rangle)$ denotes right-circular polarization.

\par Note that photon counting in the appropriate polarization
basis allows experimental determination of the Stokes parameters,
$\langle s_i \rangle$. By photon counting and using the relations
\ref{stokes1}-\ref{stokes3}, the density matrix of an arbitrary
polarization state can be reconstructed from these
data.\footnote{An extension of this technique of {\em quantum
tomography} has been demonstrated by reconstruction of two-photon
density matrices in \cite{non-max}.  This technique will be useful
in the two-photon state discussions of Sect. \ref{DFS}.} In
contrast to measurements of ${\cal V}$ with respect to some
particular orthonormal basis $\{|\lambda\rangle,\
|\overline{\lambda}\rangle\}$, this tomographic technique allows
complete determination of the polarization state.  In fact, the
visibility ${\cal V}$ with respect to any basis $|\lambda\rangle$
can be found from the density matrix if the transformation from
$|\chi_j\rangle$ to $|\lambda\rangle$ is known.

\par In the previous cases, we used an appropriately oriented waveplate
as a realization of the exchange operator, $\mathbf{R}$. However,
the choice of orientation in both cases required prior knowledge
of the eigenstates of the coupling between frequency and
polarization.  Note, however, that if we know only that the
eigenstates of the coupling are orthogonal, \textit{linear}
polarization states, then a {\em rotation} by $90^\circ$ (as
opposed to reflection about a specific axis) meets the
requirements of an exchange operator (Eqs. \ref{R1}-\ref{R2}) in
all cases. We expect that an optically active $90^\circ$ quartz
rotator acts as a quantum control element, independent of the
orientation of the quartz crystal that causes decoherence (since
the eigenstates of our birefringent quartz crystal are orthogonal
and linear, i.e., along the optic axes of the
crystal).\footnote{If the eigenstates of the environmental
coupling are left- and right-circular polarizations, a $\lambda/2$
waveplate at any orientation acts as the exchange operator ${\bf
R}$.}
    \begin{figure}
    \scalebox{.35}[.35]{\includegraphics*{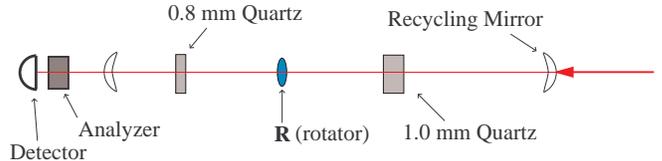}}
    \caption{\label{LinearCavity}Schematic diagram of experimental apparatus for
    the case in which the eigenstates of the polarization-frequency coupling can
    be adjusted. As before, $\mathbf{R}$ indicates the exchange element,
    an optically active
    $90^\circ$ quartz rotator in this case.  The two birefringent crystals introduce
    different phases errors whose eigenstates can be adjusted independently by
    changing their orientation.}
    \end{figure}
\par
\begin{figure}
\scalebox{0.35}[0.35]{\includegraphics{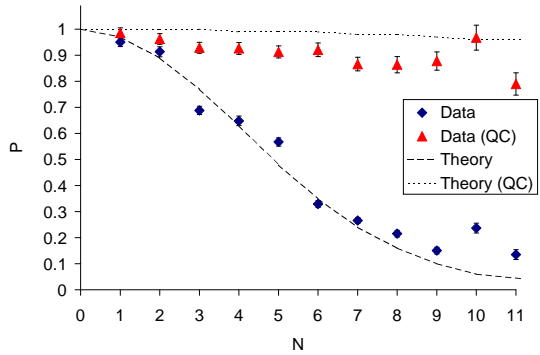}}
\caption{Theoretical and Experimental plots of $\mathcal{P}$ vs. N
for environmental eigenstates of $-10^\circ$ in a linear cavity
(see Fig. \ref{LinearCavity}). The input polarization is
$35^\circ$. Clearly, introduction of the quartz rotator results in
a preservation of photon polarization information for this choice
of environmental eigenstates.} \label{10and10}
\end{figure}

 To investigate decoherence in an ``adjustable'' linear basis, we
use a cavity with a photon passing through $1.046$ mm and $0.850$
mm birefringent quartz crystals at each cycle.  Since this
thickness of quartz results in substantially faster decoherence
than the unbalanced interferometer of the previous cases, we use a
narrow bandwidth filter of $\delta\lambda=1.5$ nm to increase the
coherence length of the photons. The exchange element, a
$90^\circ$ quartz rotator, is placed between the birefringent
crystals (see Fig. \ref{LinearCavity}). At the output, we measure
the Stokes parameters by photon counting in the appropriate bases
and calculate $\mathcal{P}$ to quantify the degree of coherence.
By orienting both birefringent crystals so that the optic axis
makes an angle of $-10^\circ$ with the horizontal, and sending
linearly polarized light at $35^\circ$ into the system (for
maximum decoherence), we see that the rotator does indeed act as a
quantum control element (Fig. \ref{10and10}). Note however, that
we need not rotate the optically active element by a corresponding
$10^\circ$, because its functionality is independent of such a
rotation. Put differently, the rotator acts as a quantum control
operator for (slow) decoherence in all linear bases.
\par For the theoretical curves in Fig. \ref{10and10}, we need not
modify the previous calculations except to take the lower
(negative) sign in the definition of $\mathbf{R}$, Eq. \ref{R2}
and include the narrower frequency spectrum. The invariance of
physical systems under rotation allows us to rotate the entire
experimental apparatus (the input polarization and quartz
crystals) by $10^\circ$ without adjusting the calculations.
\par
\begin{figure}
\scalebox{.35}[.35]{\includegraphics*{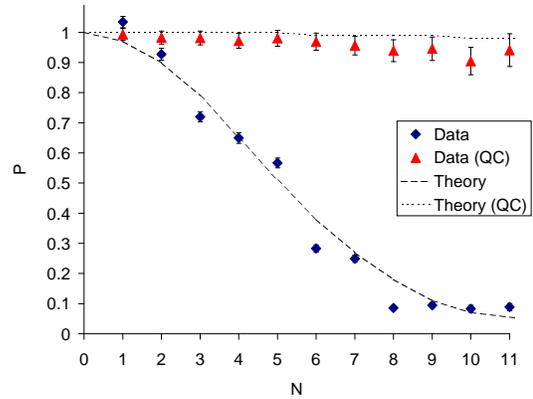}}
\caption{\label{0and10}$\mathcal{P}$ vs. N for the case in which
the strength {\em and} the eigenstates of the
polarization-frequency coupling change. In the case where we
include the quantum control element, a $90^\circ$ quartz rotator,
the degree of polarization is preserved.}
\end{figure}

 As a further demonstration of ``bang-bang'' quantum control of
decoherence, we seek a scheme in which the strength \textit{and}
the eigenstates of the polarization-frequency coupling change
between exchange operations. To reproduce these conditions, we
simply place the quartz crystals of Fig. \ref{LinearCavity} at
slightly different angles so that the eigenstates of the coupling
are not constant at each pass. The first (1.0 mm) crystal is
oriented at $10^\circ$ to the horizontal while the second (0.8 mm)
crystal is placed at $0^\circ$ degrees.  To solve for
$\mathcal{P}$ starting from the density operator representation,
we rewrite the first crystal operator as a matrix in the $H/V$
basis using the appropriate rotation matrices.  The resulting
output density matrices are calculated numerically and
$\mathcal{P}$ is extracted.  The theoretical and experimental
results are displayed in Fig. \ref{0and10}.

\begin{figure}
\scalebox{.35}[.35]{\includegraphics{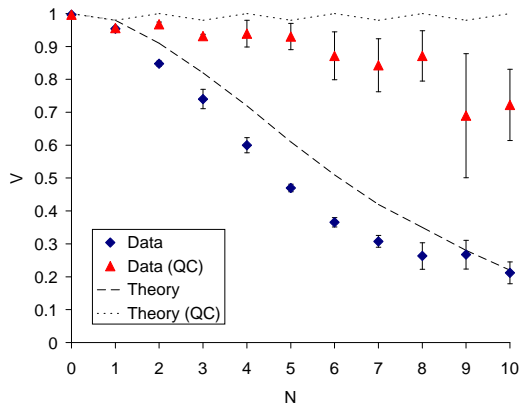}}
\caption{\label{dissipation}Experimental and theoretical results
displaying visibility vs. cycle for $T=65\%$ transmission in the
$|H\rangle$ arm of the interferometer (see Fig. \ref{michelson}).
The experimental data have large error bars since photon numbers
are quite low after 10 passes, and counting statistics (which
scale as the square root of the photon count) become significant.
As before, we expect that the imperfectly balanced interferometer
arms will reduce the visibility below the theoretical values.
Nevertheless, the principle of the quantum control for reducing
the effect of dissipation is clearly demonstrated.}
\end{figure}
\par As a final demonstration of bang-bang control, we assert that
the same technique of exchanging orthogonal polarization states
will also compensate for decoherence via {\em dissipative} (i.e.,
non-unitary) errors. Experimentally, we accomplish this task by
inserting a neutral-density ($\omega$-independent) filter into the
transmission ($|H\rangle$) arm of the {\em balanced} Michelson
interferometer of Fig. \ref{michelson}. This filter results in
$35\%$ loss (after two passes) per cycle. Denoting this operator
as ${\bf T}$, letting $T=1-0.35=0.65$, and using the input state
$\frac{1}{\sqrt{2}}(|H\rangle+|V\rangle)$, we have, after $N$
cycles,
\begin{equation}
\alpha|H\rangle+\beta|V\rangle\goesto[{\bf T}^N] \alpha
T^N|H\rangle+ \beta|V\rangle.
\end{equation}
In the case when the quantum control procedure is implemented,
${\bf T}$ and ${\bf R}$ alternate so that the state evolves in the
following steps
\begin{eqnarray}
\alpha|H\rangle+\beta|V\rangle&\goesto[{\bf T}]& \alpha
T|H\rangle+\beta |V\rangle\nonumber\\ &\goesto[{\bf R}] & \alpha
T|V\rangle+\beta|H\rangle\nonumber\\ & \goesto[{\bf T}]& \alpha
T|V\rangle+\beta T|H\rangle\nonumber\\ &\goesto[{\bf R}]&\alpha
T|H\rangle+\beta T|V\rangle
\end{eqnarray}
so that
\begin{eqnarray}
 ({\bf TR})^2(\alpha|H\rangle+\beta|V\rangle)&=&T(\alpha|H\rangle+\beta|V\rangle).
\end{eqnarray}
In this way, we prove the operator identity $({\bf TR})^2=T$ so
that, after an even number of cycles, the system evolves back to
its original state but with a reduced probability of detection.
Experimental and theoretical results for this case are displayed
in Fig. \ref{dissipation}.

\par In summary, we have shown that coupling information-carrying
(polarization) degrees of freedom to unobserved (frequency)
degrees of freedom results in a loss of polarization phase
information. We have shown in this section that an optical
realization of ``bang-bang'' control of decoherence does indeed
preserve photon polarization in the presence of such an
``environment.''  Our implementation of bang-bang control requires
some knowledge of the eigenstates of the polarization-frequency
coupling.  We have demonstrated such control when the eigenstates
of the coupling are linear and orthogonal.\footnote{In fact, in
the Poincar\'e Sphere representation of polarization (See
\cite{born_and_wolf} $\S 1.4$), it is sufficient to know only the
\textit{plane} in which the eigenstates of the coupling lie: a
$90^\circ$ rotation about the normal to that plane satisfies the
conditions of an exchange operator $\mathbf{R}$. The technique
fails, however, for a decoherence mechanism whose eigenstates
coincide with those of ${\bf R}$ (i.e., which are parallel to this
normal direction).} In addition to preserving polarization phase
information in the presence of a frequency-polarization coupling,
we have shown that such bang-bang control similarly reduces
decoherence when the strength and/or the eigenstates of the
coupling vary between exchanges.

\section{Decoherence-free subspaces of two photons}\label{DFS}

\subsection{Polarization-frequency coupling in correlated photon pairs}
\label{frequency_anticorrelation}

\par It has been shown that the antisymmetric 2-qubit
state given by (in our optical notation) $|\psi^-\rangle =
\frac{1}{\sqrt{2}}(|\chi_1\rangle|\chi_2\rangle-|\chi_2\rangle
|\chi_1\rangle)$ is immune to collective decoherence, i.e.
decoherence arising from an operation that is invariant under
qubit permutations \cite{zanardi}. That is, due to its symmetry
properties, $|\psi^-\rangle$ is a decoherence-free subspace (DFS)
of 2 qubits. Therefore, it is reasonable that we expect the
two-photon polarization state
$\frac{1}{\sqrt{2}}(|H\rangle|V\rangle - |V\rangle|H\rangle)$ to
be immune to decoherence of the type discussed in Sect.
\ref{one-photon}, which arises from coupling to frequency degrees
of freedom that are unused for information representation.

\par In this section, in order to investigate the possibility of a
DFS, we will consider polarization phase information in a
two-photon system. In non-linear optical materials, spontaneous
parametric down-conversion events can be used to produce spatially
separated, polarization-entangled photon pairs with high fringe
visibility \cite{source,ultrabright}. Here, we extend the results
of Sect. \ref{one-photon} by presenting a method for calculating
the density matrix representing the polarization of such a
two-photon state in the presence of a birefringent environment.
Experimentally, we use the source detailed in \cite{ultrabright}
which produces correlated photon pairs in the two paths labeled L
and R (See Fig. \ref{source}, p. \pageref{source}).
\par In order to keep track of frequency, spatial mode (L or R), and
polarization, we define a shorthand notation for indexing the
4$\times$4 two-photon density matrix $\rho (x)$.  The element
$|\chi_{ij}\rangle\langle\chi_{kl}|$ is defined to be
$|\chi_{i}\rangle_{L}|\chi_{j}\rangle_{R} \langle\chi_{k}|_
{L}\langle\chi_{l}|_{R}$ where the subscripts L and R refer to the
left and right photon paths respectively and $|\chi_{1}\rangle$
and $|\chi_{2}\rangle$ refer to the polarization basis states.  As
a concrete example, in the $H/V$ basis, the state
$|V\rangle_L|H\rangle_R$ is written as $|\chi_{21}\rangle$ and
$|V\rangle_L|H\rangle_R\langle H|_R\langle H|_L$ is written as
$|\chi_{21}\rangle\langle\chi_{11}|$. In this way, each of the 16
entries of $\rho$ is indexed by four numbers, ($ij,kl$), where
$i,j,k,l\in\{1,2\}$.

\par In order to satisfy energy conservation, the frequencies of photon
pairs produced in a down-conversion event must sum to the
frequency of the pump beam, and their resultant energies become
entangled (see \cite{mandel_and_wolf} $\S 22.4$). Denoting the
pump frequency by $\omega_o$ and the frequency of daughter photon
$j$ by $\omega_j$ ($j\in \{1,2\}$), we therefore require
\begin{subequations}
\begin{eqnarray}\label{anti-correlation1}
\omega_1&=&\frac{\omega_o}{2}+\epsilon\\
\omega_2&=&\frac{\omega_o}{2}-\epsilon\label{anti-correlation2}
\end{eqnarray}
\end{subequations}
so that we have $\omega_1+\omega_2 = \omega_o$ in all
cases.\footnote{We have used the notation of \cite{franson} where
such frequency entanglement is shown to cause nonlocal
cancellation of dispersion to second order in optical systems.}
This frequency ``anti-correlation'' will be important in an
experimental demonstration of a DFS.

\par Taking $c_{ij}$ to be the (in general, complex) amplitude of
the state $|\chi_{ij}\rangle$ and denoting the frequency spectrum
by $\mathrm{A}(\omega)$, we write the initial (pure) state at
$x=0$ as\footnote{We impose the normalization condition
$$\int\mathrm{d}\epsilon|A(\frac{\omega_o}{2}+\epsilon)|^{2}|A(\frac{\omega_o}{2}-\epsilon)|^{2}=1$$
so that initially $\langle\Psi|\Psi\rangle = 1$ and
$\rho^2=\rho$.}
\begin{eqnarray}
|\Psi (0)
\rangle&=&\sum_{i,j=1}^{2}c_{ij}|\chi_{ij}\rangle\nonumber\\
&&\otimes\int\mathrm{d}\epsilon\
\mathrm{A}(\frac{\omega_o}{2}+\epsilon)
\mathrm{A}({\frac{\omega_{o}}{2}}-\epsilon)|\frac{\omega_o}{2}+\epsilon\rangle|\frac{\omega_o}{2}
-\epsilon\rangle.\nonumber\\
\end{eqnarray}

\par For completeness, and to make this indexing scheme explicit, we write the
reduced $4\times4$ density matrix representing the pure
polarization state $|\Psi\rangle =
\sum_{i,j=1}^2c_{ij}|\chi_{ij}\rangle$:
\begin{equation}\label{explicit}
\rho =|\Psi\rangle\langle\Psi|=
\begin{pmatrix}
|c_{11}|^2 &c_{11}c^\ast_{21}&c_{11}c^\ast_{12}&c_{11}c_{22}^\ast
 \cr c_{21}c^\ast_{11}& |c_{21}|^2 &c_{21}c^\ast_{12}&c_{21}c_{22}^\ast
 \cr c_{12}c^\ast_{11}&c_{12}c^\ast_{21}& |c_{12}|^2 &c_{12}c_{22}^\ast
 \cr c_{22}c^\ast_{11}&c_{22}c^\ast_{21}&c_{22}c^\ast_{12}& |c_{22}|^2
\end{pmatrix}.
\end{equation}
Note that there are six independent parameters corresponding to
the four complex amplitudes of the states $|\chi_{ij}\rangle$
constrained by the normalization condition $\sum|c_{ij}|^2=1$ and
the freedom to choose the global (overall) phase.\footnote{For a
partially mixed state, there are 15 independent parameters
corresponding to 4 real diagonal elements reduced to 3 by the
normalization constraint, and 12 complex amplitudes in the upper
right entries.}
\par Introducing the shorthand notation
$\omega^{\pm} = \omega ^\pm (\epsilon) = \frac{\omega_o}{2}
\pm\epsilon$ and $\tilde{\omega}^\pm = \tilde{\omega}^\pm
(\epsilon) =\frac{\omega_o}{2} \pm\tilde{\epsilon}$, the density
operator for the initial state, including frequency modes, can be
written as
\begin{eqnarray}
\rho_\omega(0) &= &|\Psi(0)\rangle\langle\Psi(0)| \nonumber\\
&=&\sum_{i,j,k,l} c_{ij}c^{\ast}_{kl} |\chi_{ij}\rangle
\langle\chi_{kl}| \nonumber\\ &&{} \otimes\iint\mathrm{d}\epsilon\
\mathrm{d} \tilde{\epsilon} \ \left\{
\mathrm{A}(\omega^+)\mathrm{A} (\omega^-) \mathrm{A}^{\ast}
(\tilde{\omega}^+) \mathrm{A}^{\ast} (\tilde{\omega}^-)
\right.\nonumber
\\ && \left.\ \times\ |\omega^+ \rangle| \omega^- \rangle\langle
\tilde{\omega} ^+|\langle \tilde{\omega}^-| \right\}
\label{rho_correlated}
\end{eqnarray}
where we must be careful to treat $\omega^\pm$ and
$\tilde{\omega}^\pm$ as functions of the variables of integration
$\epsilon$ and $\tilde{\epsilon}$.
\par In order to represent a birefringent crystal in each path, we
define the operator
\begin{equation}\mathbf{U}(x_L,x_R)=\mathbf{U}_L(x_L)\otimes\mathbf{U}_R(x_R)
\label{UL_UR}\end{equation} which depends on the distance, $x_L$
($x_R$), traveled through such a crystal in the L (R) photon path.
It should be noted that the $\otimes$ in Eq. \ref{rho_correlated}
represents a tensor product between the Hilbert spaces of photon
polarization and photon frequency, whereas the $\otimes$ in Eq.
\ref{UL_UR} represents the tensor product of photon spatial modes
L and R. In analogy with Sect. \ref{review}, $\mathbf{U}_L(x_L)$
associates a frequency-dependent phase, $\varphi_j(\omega,x_L)$,
to each polarization state $j \in \{1,2\}$ in the path labeled L:
\begin{equation}
\mathbf{U}_L(x_L)
|\chi_j\rangle_L=\mathrm{e}^{i\varphi_j(\omega,x_L)}|\chi_j\rangle_L.
\end{equation}
A similar relation holds for the path labeled R.  The functions
$\varphi_j(\omega,x_{L/R})$ characterize the phase shift induced
in either path (L or R).  In this way, we require that the
coupling between frequency and polarization share the same
dependence on the parameters $x$ and $\omega$ in either photon
path. This requirement is reasonable since we wish to explore the
robustness of the antisymmetric state against {\em collective}
decoherence.

\par Making a frequency-insensitive measurement of polarization after the
photon in path L (R) has traveled a distance $x_L$ ($x_R$), we
have
\begin{widetext}
\begin{equation}\label{rho_formal}
\rho(x_L,x_R)=
\iint\mathrm{d}\omega_3\mathrm{d}\omega_4\langle\omega_3|\langle\omega_4
|\mathbf{U}(x_L,x_R)\rho_\omega(0)\mathbf{U^{\dag}}(x_L,x_R)|\omega_3\rangle|\omega_4\rangle
\end{equation}
which gives
\begin{eqnarray}\label{rho2}
\rho(x_L,x_R)&=&
\sum_{i,j,k,l}c_{ij}c_{kl}^{\ast}|\chi_{ij}\rangle\langle\chi_{kl}|\int\mathrm{d}
\epsilon\
|A(\omega^{+})|^{2}|A(\omega^{-})|^{2}e^{i\left(\varphi_{i}(\omega^{+},x_L)+
\varphi_{j}(\omega^{-},x_R)-\varphi_{k}(\omega^{+},x_L)-\varphi_{l}(\omega^{-},x_R)\right)}.
\end{eqnarray}
\end{widetext}
\par Eq. \ref{rho2} gives a prescription for calculating the
density matrix representing polarization degrees of freedom in the
presence of non-dissipative ``phase errors.''  As in Sect.
\ref{bang-bang}, we expect to observe decoherence effects in the
off-diagonal elements of $\rho$.  Note that for the diagonal
elements of $\rho$, indexed by $i=k,j=l$ (see Eq. \ref{explicit}),
the argument of the exponential is identically zero so that the
{\em diagonal} elements do not change.  Of course, this result
holds only for the case of a unitary frequency-polarization
coupling in this particular orientation (i.e., with eigenstates
$|H\rangle$ and $|V\rangle$).

\par Let us now choose the functions
$\varphi_j(\omega,x)$ to be identical to those of Eq.
\ref{quartz_phi}.  In other words, we choose to realize a unitary
polarization-frequency coupling by placing a birefringent crystal
across both photon paths with e- and o-axes aligned with
$|\chi_1\rangle$ and $|\chi_2\rangle$.\footnote{We can realize
this situation experimentally either by placing the same crystal
across both paths, or equivalently by placing ``identical''
crystals in each path and at the same orientation.  For future
convenience, we choose the second option.} Letting $\mathcal{L}$
denote the thickness of the crystal, the phase {\em difference}
between these functions evaluated at $x_L=x_R=\mathcal{L}$ is
given by
\begin{equation}\label{phase_before_modification}
\varphi_2(\omega,\mathcal{L})-
\varphi_1(\omega,\mathcal{L})=\omega\frac{\mathcal{L}\Delta n}{c}.
\end{equation}
\par In Appendix \ref{density_matrices}, the
$4\times4$ density matrix representation is written for the
evolution of the Bell states $|\phi^\pm\rangle=
\frac{1}{\sqrt{2}}(|HH\rangle\pm|VV\rangle)$ and
$|\psi^\pm\rangle=\frac{1}{\sqrt{2}}(|HV\rangle\pm|VH\rangle)$.
From these, we see that the states $|\psi^\pm\rangle$ become
``incoherent'' in the presence of such a frequency-polarization
coupling. Here, we do not make a quantitative measure of coherence
(or mixture) but simply observe that the off-diagonal elements of
the density matrix approach zero for $\mathcal{L}$ larger than the
coherence length of the down-converted photons, given by
$\sim\frac{c}{\delta\omega}$ where $\delta\omega$ is the width of
the frequency spectrum $|A(\omega)|^2$. Therefore, these states
lose phase coherence information in the presence of
polarization-frequency coupling of this type: identical
birefringent crystals in the same orientation in both photon
paths.

\par On the other hand, the states $|\phi^\pm\rangle$ do \textit{not} undergo
decoherence for this particular coupling: they retain a definite
phase relationship in the off-diagonal matrix elements. We see
from the density matrix representation that the state
$\frac{1}{\sqrt{2}}(|HH\rangle\pm|VV\rangle)$ at the input looks
like $\frac{1}{\sqrt{2}}(|HH\rangle\pm
e^{i\frac{\omega_o}{2}\frac{\mathcal{L}\Delta n}{c}} |VV\rangle)$
at the analyzer.\footnote{The phase factor arises from the
\textit{pump} frequency $\omega_o$ which we have assumed to be
effectively monochromatic so that only a definite phase shift, and
not decoherence, occurs. Including a frequency spectrum of finite
width for the pump would result in additional decoherence effects
on a different length scale, the coherence length of the pump
beam.} For this particular arrangement, a birefringent crystal
with axes along $H$ and $V$, the states $|HH\rangle\pm|VV\rangle$
\textit{do not decohere}: there is no decline in magnitude of the
off-diagonal elements of the density matrix.

\par Note, however, that neither of the states $|\phi^\pm\rangle$ can
be considered a decoherence-free subspace (DFS). In general, with
respect to another (possibly elliptical) basis $\{|\lambda\rangle
|\overline{\lambda}\rangle\}$,
\begin{equation}
\frac{1} {\sqrt{2}} (|HH\rangle\pm|VV\rangle \neq\frac{1}
{\sqrt{2}} (|\lambda\lambda\rangle\pm| \overline{\lambda}
\overline{\lambda}\rangle.
\end{equation}
In particular, using the basis of left- and right-circular
polarization, it can be shown that
$\frac{1}{\sqrt{2}}(|HH\rangle+|VV\rangle) =
\frac{1}{\sqrt{2}}(|LR\rangle+|RL\rangle)$.  So, just as the state
$\frac{1}{\sqrt{2}}(|HV\rangle+|VH\rangle)$ loses phase
information in a birefringent crystal whose eigenmodes (for
photons in either path) are $|H\rangle$ and $|V\rangle$, the state
$\frac{1}{\sqrt{2}}(|LR\rangle+|RL\rangle)$ loses phase
information in a crystal whose eigenmodes are $|L\rangle$ and
$|R\rangle$.  On these grounds, the only candidate for a DFS is
the singlet state $|\psi^-\rangle =
\frac{1}{\sqrt{2}}(|HV\rangle-|VH\rangle)$, since, in analogy with
the spin-$\frac{1}{2}$ singlet state
$\frac{1}{\sqrt{2}}(|\uparrow\downarrow\rangle -
|\downarrow\uparrow\rangle)$, this state is rotationally invariant
and looks the same when written in any basis (see \cite{sakurai},
$\S 3.9$). However, we have already shown that the state
$|\psi^-\rangle$ loses phase information in this system.

\subsection{Recovering a decoherence-free subspace}\label{phase-anticorrelation}

\par The apparent contradiction between the prediction of a DFS
and the contrary result in the previous section arises as a result
of the frequency entanglement (anti-correlation) in down-converted
photon pairs. We expected to observe a DFS of two-photon states
subjected to {\em identical} phase errors. Frequency
anti-correlation breaks this qubit permutation symmetry since the
phase errors are frequency-dependent. There is, however, a simple
experimental alteration which compensates for this frequency
anti-correlation and allows us to observe a DFS as was originally
expected.

\par Global phase shifts are never observable: only the phase
\textit{difference} induced between eigenstates of the coupling
operator enters into the calculations. As a result of frequency
anti-correlation, the phase shift experienced by a photon in path
L with frequency $\frac{\omega_o}{2}+\epsilon$ corresponds to that
of an (identically polarized) photon of frequency
$\frac{\omega_o}{2}-\epsilon$ in path R.\footnote{This equivalence
holds only in the limit that we neglect dispersion effects in the
crystals so that $\Delta n$ of Eq. \ref{phase_before_modification}
is independent of $\omega$.  This is true to a very good
approximation over the 5 nm bandwidth of the photons in our
experiment.} We have already seen that cancellation of phase terms
in the exponential of Eq. \ref{rho2} accounts for the coherence
properties of a particular polarization state. In particular, an
off-diagonal matrix element goes to zero if not all $\epsilon$
terms cancel out, and the Fourier integral does not reduce to
unity. Since frequency anti-correlation directly affects the sign
of these phase terms, it also affects their cancellation or
non-cancellation and consequently alters the predicted
decoherence-free nature of the singlet state.

\par By inducing a second anti-correlation, which we shall term
``path anti-correlation,'' we recover the proposed DFS. This
operation consists of forcing the phase difference in path L to be
the negative of the corresponding phase difference in path R (for
identical frequencies).  By twice reversing the sign of
corresponding phase shifts (once with frequency anti-correlation
and a second time with path anti-correlation) we effectively
produce identical phase errors in each path, and restore the
permutation symmetry of the polarization-frequency coupling.

\par We are now faced with the experimental task of forcing phase shifts for
corresponding polarizations to be of opposite sign in opposite
photon paths. Experimentally, we may simply rotate the crystal in
one arm by $90^\circ$. This operation exchanges $n_e$ and $n_o$
and reverses the sign of $\Delta n$ in Eq.
\ref{phase_before_modification}.\footnote{In general, we can
produce a similar frequency-polarization coupling with any
elliptical eigenmodes $|\lambda\rangle,
|\overline{\lambda}\rangle$ by rotating these eigenstates into
$|H\rangle$ and $|V\rangle$ by passive optical elements (a unitary
transformation), introducing a birefringent crystal aligned along
$H$ and $V$, and subsequently inverting the polarization
transformation with more optical elements.  Also, in this more
general case, phase anti-correlation is realized by rotating the
birefringent crystal in one arm by $90^\circ$.}

\par To this end, we introduce a new (but familiar looking)
operator defined by
\begin{equation}
\tilde{\mathbf{U}}(x_L,x_R) = \mathbf{U}^\perp_L(x_L) \otimes
\mathbf{U}_R(x_R).
\end{equation}
$\mathbf{U}^\perp_L(x_L)$ represents the crystal in path L which
has been rotated and is now characterized by the phase function
$\tilde{\varphi}_j$ so that
\begin{subequations}
\begin{eqnarray}
\mathbf{U}^\perp_L|\chi_j\rangle_L&=&e^{i
\tilde{\varphi}_j(\omega,x_L)} |\chi_j\rangle_L \\
\mathbf{U}_R|\chi_j\rangle_R&=&e^{i \varphi_j(\omega,x_R)}
|\chi_j\rangle_R .
\end{eqnarray}
\end{subequations}
The phase functions $\tilde{\varphi}_1$ and $\tilde{\varphi}_2$ in
path L are related to the phase shifts $\varphi_1$ and $\varphi_2$
in path R by
\begin{equation}\label{phase_after_modification}
\tilde{\varphi}_2(\omega, \mathcal{L}) - \tilde{\varphi}_1(\omega,
\mathcal{L})= {\varphi}_1(\omega, \mathcal{L}) - \varphi_2(\omega,
\mathcal{L}).
\end{equation}
(compare Eq. \ref{phase_before_modification}). Keeping in mind
that photon states with spatial mode L are indexed by $i$ and $k$,
we compute the analog of Eq. \ref{rho2}.
\begin{eqnarray}\label{rho2_modified}
&&\rho(x_L,x_R)= \nonumber \\
&&\sum_{i,j,k,l}c_{ij}c_{kl}^{\ast}|\chi_{ij}\rangle\langle\chi_{kl}|
\int\mathrm{d} \epsilon\
\left\{|A(\omega^{+})|^{2}|A(\omega^{-})|^{2}\right.\nonumber \\
&&\times \left.e^{i\left(\tilde{\varphi_{i}} (\omega^{+},x_L)+
\varphi_{j} (\omega^{-},x_R)- \tilde{ \varphi_{k}}
(\omega^{+},x_L)-\varphi_{l} (\omega^{-},x_R)\right)}\right\}.
\nonumber\\
\end{eqnarray}

\par The $4\times 4$ matrices representing the evolution of the four
Bell states are displayed in Appendix \ref{density_matrices}. From
these, we see that the states $|\phi^\pm\rangle$ at the input
become mixed, or incoherent, in that the magnitude of the
off-diagonal elements approaches zero. On the other hand, the
states $|\psi^\pm\rangle$ are possibly transformed at the pump
frequency, but \textit{do not decohere}. In particular, for the
initial state $\rho(0)=|\psi^-\rangle\langle\psi^-|$, the output
state is
\begin{equation}
\rho(0)\rightarrow\frac{1}{2}
\begin{pmatrix}
     0 & 0 & 0 & 0
 \cr 0 & 1 & - e^{-i\omega_o\frac{\mathcal{L}\Delta n}{c}} & 0
 \cr 0 & - e^{i\omega_o\frac{\mathcal{L}\Delta n}{c}} & 1 & 0
 \cr 0 & 0 & 0 & 0
\end{pmatrix}.
\end{equation}
The state
$|\psi^-\rangle=\frac{1}{\sqrt{2}}(|HV\rangle-|VH\rangle)$ at the
input looks like $\frac{1}{\sqrt{2}}(|HV\rangle -e^{i
\frac{\omega_o}{2}\frac{\mathcal{L}\Delta n}{c}}|VH\rangle)$ at
the output.\footnote{In our experiment, we adjusted ${\cal L}$ (by
slightly tilting one of the quartz elements) such that this phase
factor is $0$ (modulo $2\pi$).}
\par Recall now that the singlet state, $|\psi^-\rangle$ can be written
as $\frac{1}{\sqrt{2}}(|\lambda \overline{\lambda} \rangle -
|\overline{\lambda} \lambda \rangle)$ where $|\lambda\rangle$ and
$|\overline{\lambda}\rangle$ are the (in general, elliptical)
eigenmodes of \textit{any} birefringent environment. We conclude,
therefore, that the state $|\psi^-\rangle$ is generally immune to
decoherence of this type (i.e., collective decoherence, produced
here with perpendicular birefringent elements in each arm).  By
imposing an experimental requirement which restores the symmetry
of the coupling with respect to photon path permutation, we have
recovered the singlet state as a DFS with respect to collective
decoherence.

\begin{figure}
\scalebox{.35}[.35]{\includegraphics{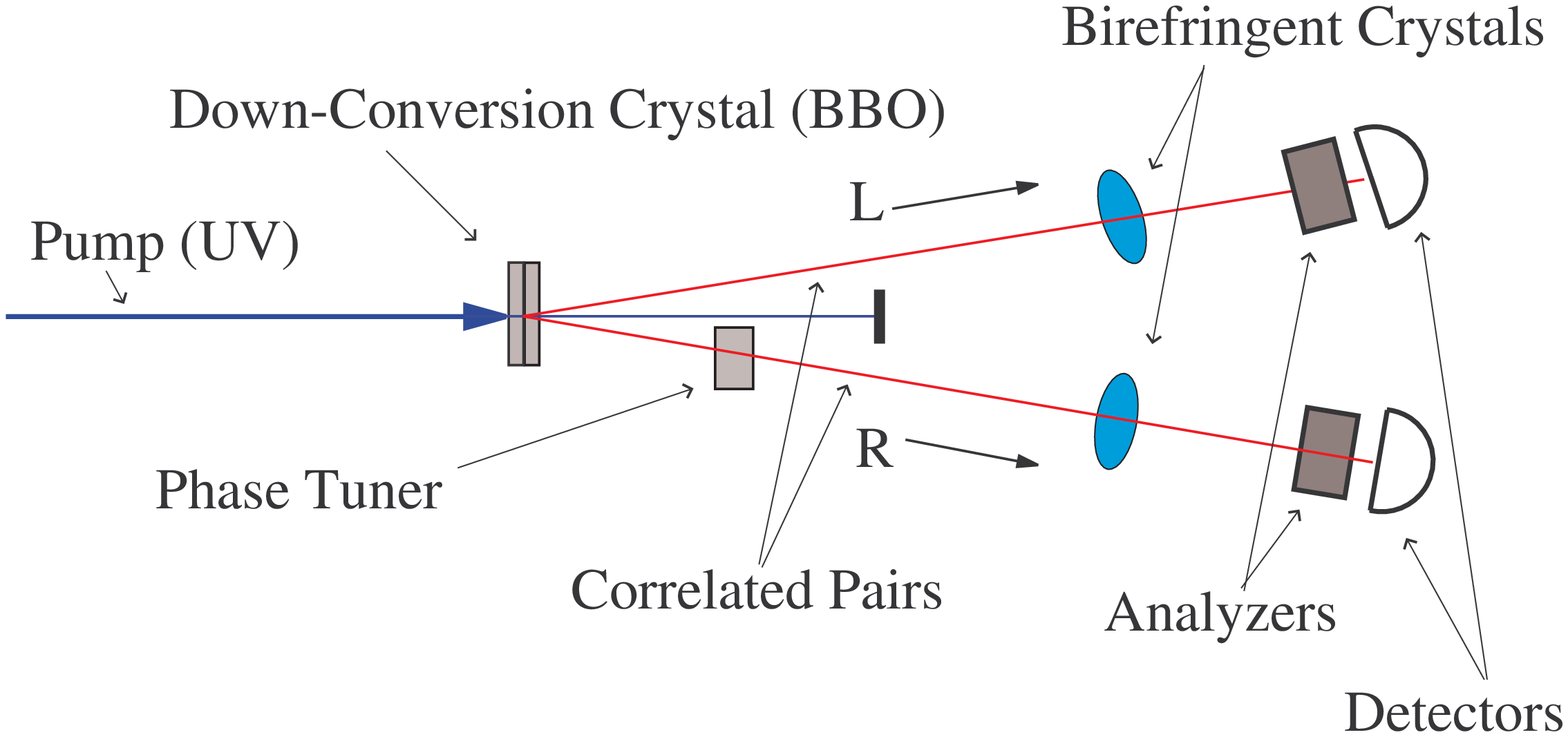}}
\caption{\label{source}Schematic diagram of an experimental
apparatus for producing and analyzing correlated photon pairs. The
phase tuner is used to produce the desired input state.  The
birefringent crystals have identical thickness, but can be
oriented independently.  For the experimental results, the crystal
in path L was rotated by $\theta_L=-17^\circ$, and the crystal in
path R was rotated by $\theta_R=-107^\circ$.  The birefringent
crystals give an effective optical path difference between $H$ and
$V$ of $\sim140\lambda_o$, where $\lambda_o$ is the central
wavelength of the down-converted photons (702 nm, in our case).}
\end{figure}

\par The technique of quantum tomography allows
experimental reconstruction of two-photon polarization state
density matrices by an appropriate choice of polarization
measurements (on members of an ensemble) \cite{non-max}. With the
experimental setup shown in Fig. \ref{source} and using this
technique, we compare the theoretical predictions with
experimentally observed density matrices. Details of the
experiment and results will be reported elsewhere
\cite{DFS_paper}.  Here, we simply present the theroetical
predictions.
\par We seek to quantify decoherence properties in a single
parameter. To this end, we use the Fidelity of the transmission
process, defined here as $F = Tr(\rho_{out} \rho_{in})$ where
$\rho_{in}$ ($\rho_{out}$) represents the polarization density
matrix at the input (output) of the system \cite{shumacher}.  For
a mixed input state, we must use $F =
\left[Tr(\sqrt{\sqrt{\rho_{in}}\rho_{out}\sqrt{\rho_{in}}}\right]^2$
\cite{mixedFid}. A Fidelity of $1$ indicates a decoherence-free
process. Numerical evaluation of Eq. \ref{rho2_modified} allows us
to calculate the Fidelity of the transmission process
for coupling eigenstates rotated by
an arbitrary angle.  Such an analysis reveals a Fidelity of 1 for
$|\psi^-\rangle$ in a birefringent environment with arbitray
eigenstates, confirming the prediction that the singlet state is a
decoherence-free subspace.

\par
\begin{figure}
    \begin{tabular}{cc}
    \textbf{Input}&\textbf{Output} \\
    \raisebox{10ex}{$\frac{1}{\sqrt{2}}(|HH\rangle+|VV\rangle)$\ \ }
    & \scalebox{0.3}[0.3]{\includegraphics{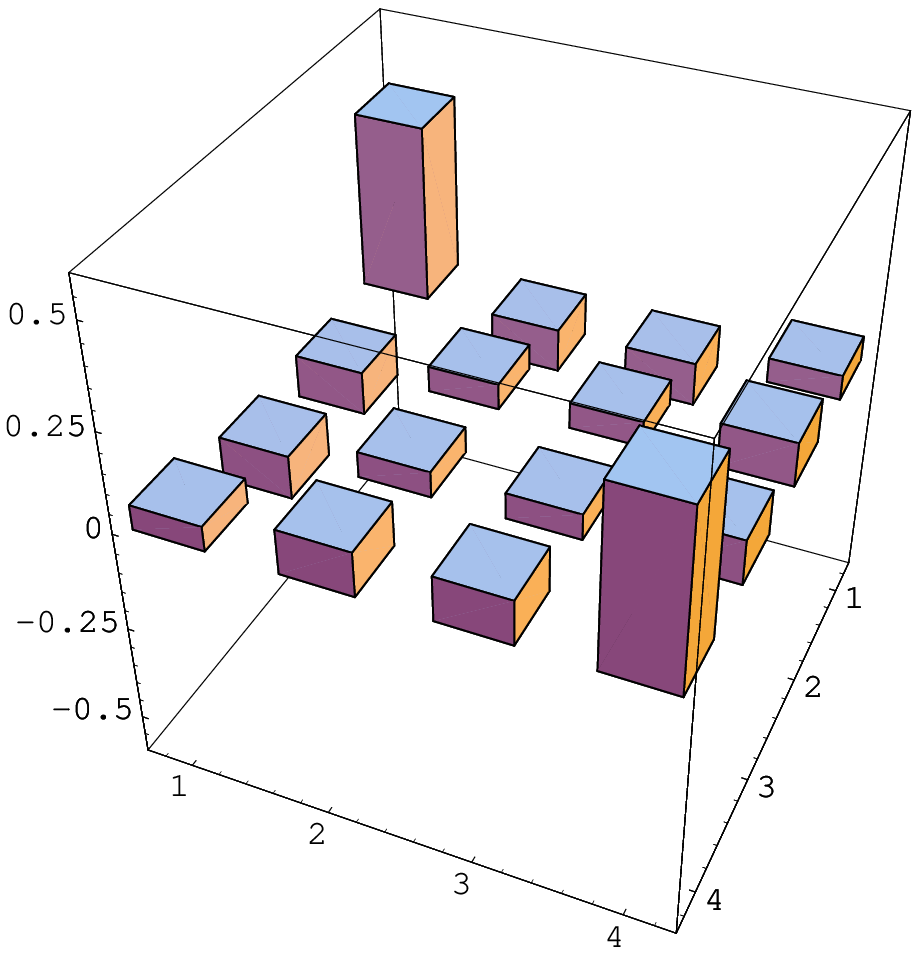}}\\
    \raisebox{10ex}{$\frac{1}{\sqrt{2}}(|HH\rangle-|VV\rangle)$\ \ }
    & \scalebox{0.3}[0.3]{\includegraphics{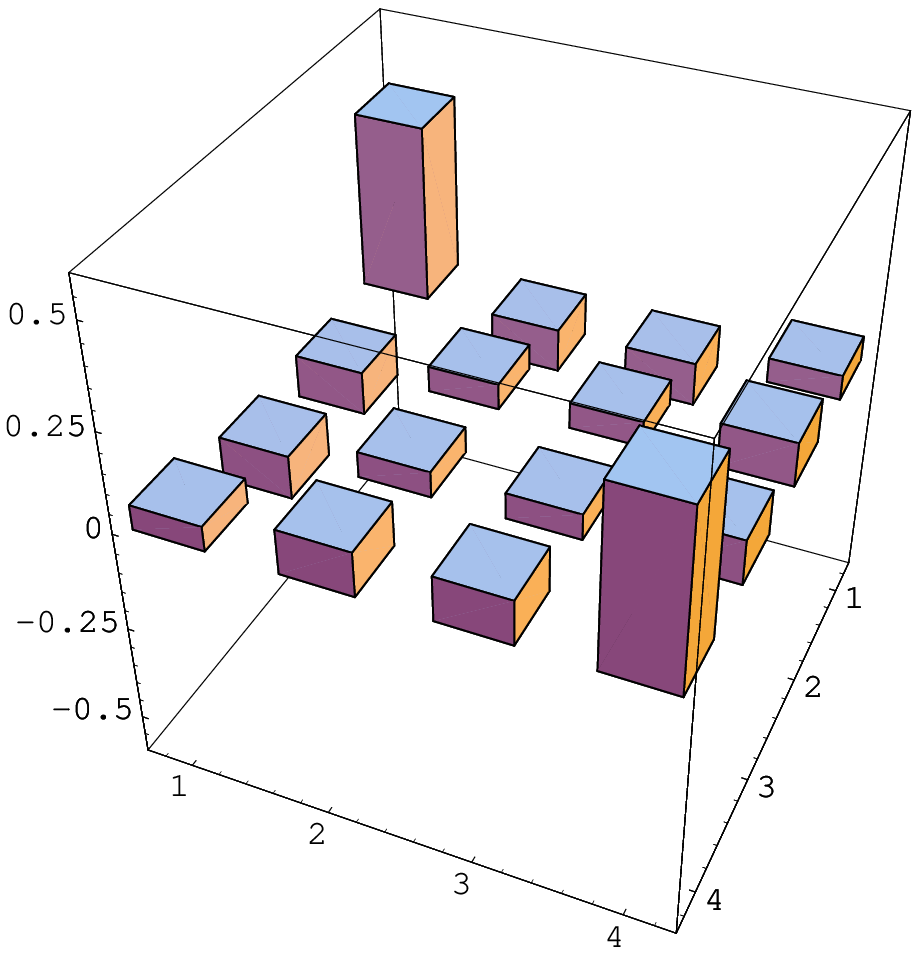}}\\
    \raisebox{10ex}{$\frac{1}{\sqrt{2}}(|HV\rangle+|VH\rangle)$\ \ }
    & \scalebox{0.3}[0.3]{\includegraphics{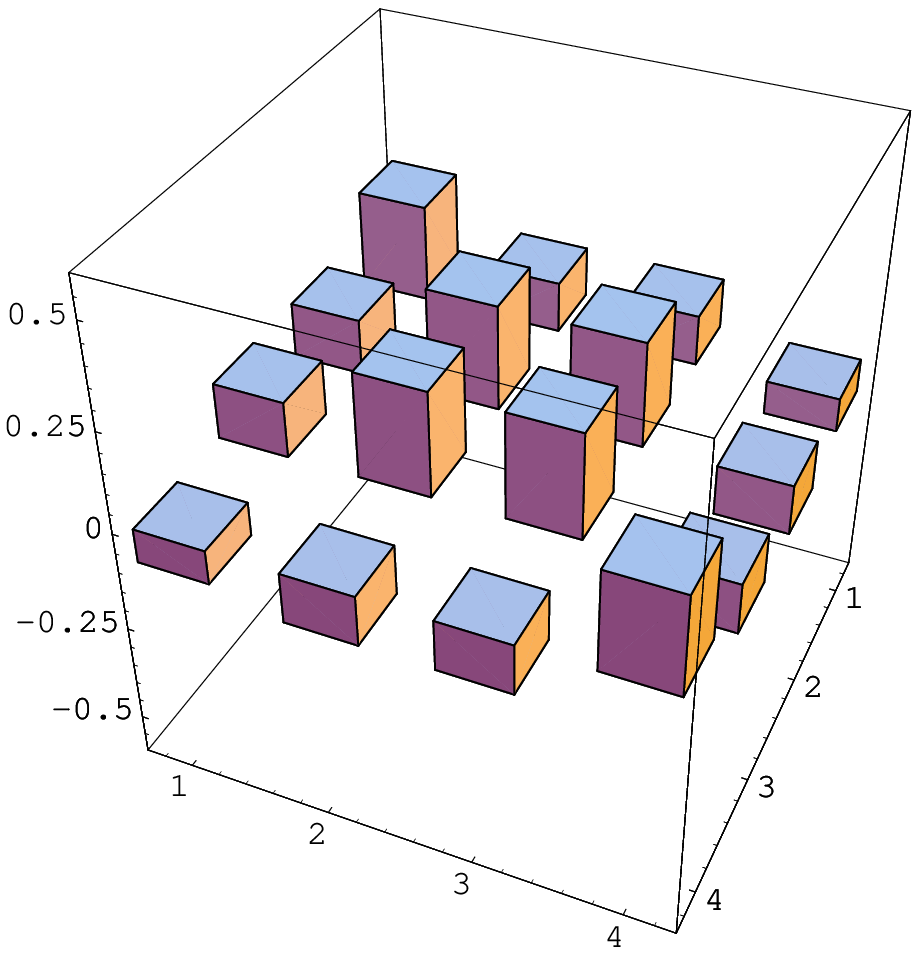}} \\
    \raisebox{10ex}{$\frac{1}{\sqrt{2}}(|HV\rangle-|VH\rangle)$\ \ }
    & \scalebox{0.3}[0.3]{\includegraphics{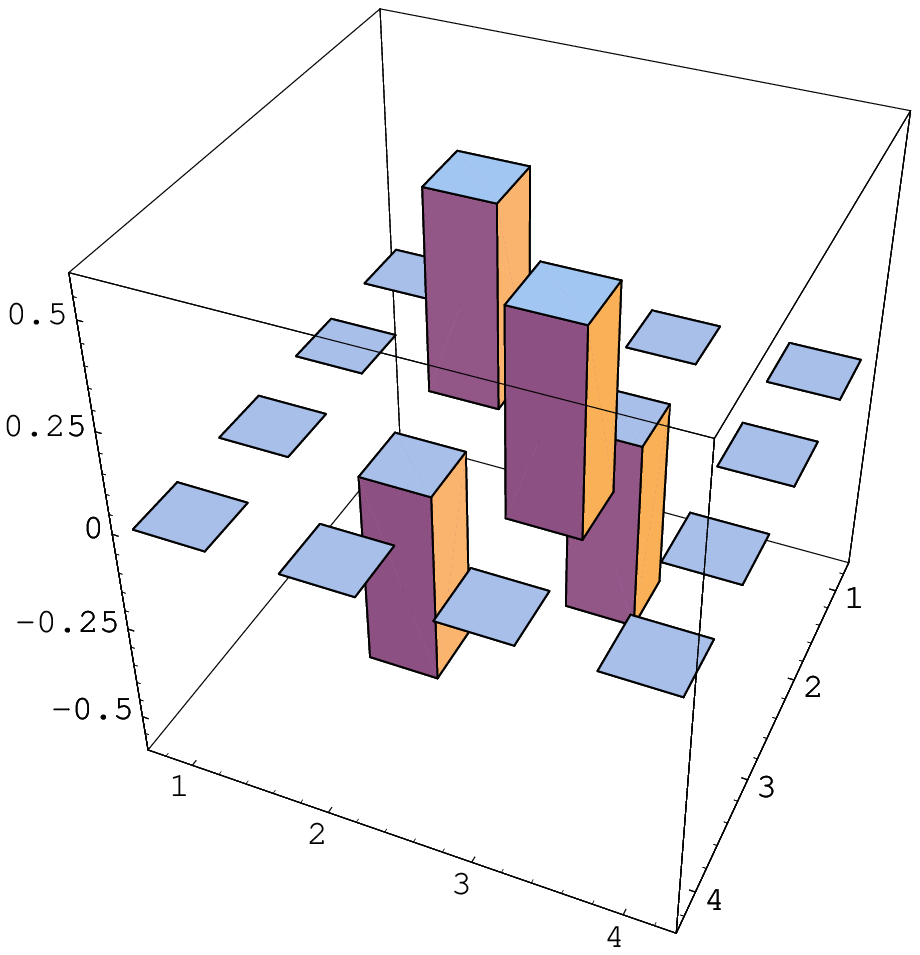}}\\
    &\\
    \end{tabular}
    \caption{\label{DFS_phi}Real part of polarization state
    density matrices for Bell states at the input of the
    experimental setup of Fig. \ref{source} as predicted by Eq.
    \ref{rho2_modified}.
    Imaginary parts are zero for all elements.}
    \end{figure}

 In Fig. \ref{DFS_phi}, we present a graphical
representation of the density matrices for each of the four Bell
States passing through birefringent crystals in each photon path
as calculated from Eq. \ref{rho2_modified}.

We see that the singlet state does indeed maintain full coherence
in the presence of a unitary frequency-polarization coupling.
Since polarization degrees of freedom are analyzed and frequency
degrees of freedom are unobserved, such a result demonstrates the
robustness of a decoherence-free subspace of two qubits against
non-dissipative errors arising from coupling with unutilized
(environmental) degrees of freedom.
\section{Summary and Conclusions}

\par In summary, we have developed a prescription for
calculating density matrices representing one- and two-photon
states in the presence of unitary frequency-polarization coupling.
With these methods, we have investigated decoherence in quantum
systems arising from coupling to unobserved (environmental)
degrees of freedom. By inducing frequency-dependent phase errors
in photon polarization states and tracing over frequency degrees
of freedom, we observe information loss as a particular form of
decoherence.

\par In the one photon case, we applied the results
of \cite{bang-bang} and established the feasibility and utility of
``bang-bang'' quantum control of decoherence in an optical system.
By periodically exchanging the eigenstates of the
``environmental'' coupling, photon coherence (polarization
information) is preserved.  We observed this effect in the case in
which the {\em strength} of the coupling is varied (the
``slow-flipping'' case); the case in which the {\em basis} of the
coupling is varied; and the case in which both the basis {\em and}
the strength of the coupling are varied simultaneously.
Additionally, we demonstrated the utility of bang-bang control in
the presence of a non-unitary (dissipative) coupling operation. In
all cases, we see a qualitative reduction in decoherence for the
case in which the quantum control procedure is implemented.

\par In the two-photon case, we applied general results
concerning the existence of decoherence-free subspaces in
\cite{zanardi,lidar} to the case of polarization entangled photon
pairs produced in a down-conversion scheme.  We found that energy
conservation in the down-conversion process gives rise to
frequency entanglement which alters the assumptions underlying the
prediction of a DFS.  In particular, frequency anti-correlation
breaks the permutation symmetry of the environmental coupling
between opposite photon paths by anti-correlating their respective
phase errors. We found that modifying the experimental apparatus
to restore the symmetry of the phase errors allows us to produce a
truly collective decoherence process, and consequently we
recovered the singlet state as a DFS.

\par In conclusion, we have investigated two procedures for
reducing the decoherence of an optical qubit subjected to
non-dissipative phase errors and non-unitary dissipative errors.
The first procedure, ``bang-bang'' quantum control, preserves
polarization information in a single-photon by rapid, periodic
control operations which effectively average out decoherence
effects.  The second procedure requires entanglement between two
photons so that the resulting state exhibits symmetry properties
which are immune to collective decoherence. These cases together
demonstrate the feasibility of both active and passive protection
of coherence properties in an optical qubit.
\appendix
\section{Mathematical Arguments}\label{properties}
\subsection{Eigenvalues of
$\mathbf{U}(x)$}\label{eigenvalues}
    \par Unitarity restricts the possible eigenvalues,
    $U_j(\omega,x)$, of ${\bf U}(x)$:
    \begin{eqnarray}
    1& =&\langle\chi_j|\chi_j\rangle \nonumber \\
     &=&\langle\chi_j|\mathbf{U^\dag}(x)\mathbf{U}(x)
    |\chi_j\rangle \nonumber
    \\&=&\langle\chi_j|U_j^\ast(\omega,x) U_j(\omega,x)|\chi_j\rangle
    \nonumber
    \\ &=&| U_j(\omega,x)|^2.
    \end{eqnarray}
    Since $U_j(\omega,x)$ is complex with modulus 1, we may write
    \begin{equation}
    U_j(\omega,x)= e^{i \varphi_j(\omega,x)}
    \end{equation}
    where $\varphi_j$ is a real-valued function of $\omega$ and $x$,
    and the subscript refers to polarization mode $j$.
\subsection{Operator Relations with $\mathbf{R}$ and
$\mathcal{U}$}\label{op_identities}
    \subsubsection{Proof of Equations \ref{identity1},
    \ref{identity2}}\label{proof1}
    Since $\mathcal{U}$ is a linear operator, we need only show
    that Eqs. \ref{identity1} and \ref{identity2} hold for the basis states,
    $|\chi_1,\omega\rangle$ and $|\chi_2,\omega\rangle$.  From the
    definitions of $\mathcal{U}$ and $\mathbf{R}$ (Eqs. \ref{R1}-\ref{scriptU2}), we have
    \begin{subequations}
    \begin{eqnarray}
    \mathbf{R}\mathcal{U}\mathbf{R}\mathcal{U}|\chi_1,\omega\rangle &=&
    \mathbf{R}\mathcal{U}\mathbf{R}|\chi_1,\omega\rangle\nonumber\\
    &=&\mathbf{R}\mathcal{U}|\chi_2,\omega\rangle\nonumber\\
    &=&e^{i\varphi(\omega)}\mathbf{R}|\chi_2,\omega\rangle\nonumber\\
    &=&\pm e^{i\varphi(\omega)}|\chi_1,\omega\rangle.
    \end{eqnarray}
    Similarly,
    \begin{eqnarray}
    \mathbf{R}\mathcal{U}\mathbf{R}\mathcal{U}|\chi_2,\omega\rangle&=&
    e^{i\varphi(\omega)}\mathbf{R}\mathcal{U}\mathbf{R}|\chi_2,\omega\rangle\nonumber\\
    &=&\pm e^{i\varphi(\omega)}\mathbf{R}\mathcal{U}|\chi_1,\omega\rangle\nonumber\\
    &=&\pm e^{i\varphi(\omega)}\mathbf{R}|\chi_1,\omega\rangle\nonumber\\
    &=&\pm e^{i\varphi(\omega)}|\chi_2,\omega\rangle.
    \end{eqnarray}
    \end{subequations}
    Thus we have, $\left(\mathbf{R}\mathcal{U}\right)^2 = \pm
    e^{i\varphi(\omega)}$ where the upper (lower) sign occurs when
    $\mathbf{R}$ represents a polarization reflection (rotation).

    It follows immediately that
    \begin{equation} \left(\mathcal{U}^\dag\mathbf{R}^\dag\right)^2
    = \pm e^{-i\varphi(\omega)}.
    \end{equation}
    Since equations \ref{identity1} and \ref{identity2} hold for
    the basis states they hold for all linear combinations of basis
    states, which is to say, for all states.

    \subsubsection{Derivation of Equations \ref{relation1}-\ref{relation4}}\label{proof2}
    \par All four relations follow easily from the definitions of
    $\mathbf{R}$ and $\mathcal{U}_j$ (Eqs. \ref{R1}-\ref{R2} and \ref{scriptU11}-\ref{scriptU22}).
    Equation \ref{relation1} is trivial, since by definition, both operators
    $\mathcal{U}_1$ and $\mathcal{U}_2$ act as the identity on $|\chi_1\rangle$.
    For the other three, we apply the definitions in succession.
    \begin{eqnarray}
    \mathcal{U}_2\mathcal{U}_1|\chi_1\rangle&=&|\chi_1\rangle\\
    \mathcal{U}_2\mathcal{U}_1|\chi_2\rangle&=&\mathcal{U}_2e^{i\varphi_1(\omega)}|\chi_2\rangle\nonumber\\
    & = &e^{i\left(\varphi_1(\omega)+\varphi_2(\omega)\right)}|\chi_2\rangle\\
    \mathbf{R}\mathcal{U}_2\mathbf{R}\mathcal{U}_1|\chi_1\rangle
    &=&\mathbf{R}\mathcal{U}_2\mathbf{R}|\chi_1\rangle\nonumber\\
    &=&\mathbf{R}\mathcal{U}_2|\chi_2\rangle\nonumber\\
    &=&\mathbf{R}e^{i\varphi_2(\omega)}|\chi_2\rangle\nonumber\\
    &=&\pm e^{i\varphi_2(\omega)}\left(|\chi_1\rangle\right)\nonumber\\
    &\rightarrow&|\chi_1\rangle\label{a7}\\
    \mathbf{R}\mathcal{U}_2\mathbf{R}\mathcal{U}_1|\chi_2\rangle
    &=&\mathbf{R}\mathcal{U}_2\mathbf{R}e^{i\varphi_1(\omega)}|\chi_2\rangle\nonumber\\
    &=&\pm\mathbf{R}\mathcal{U}_2e^{i\varphi_1(\omega)}|\chi_1\rangle\nonumber\\
    &=&\pm\mathbf{R}e^{i\varphi_1(\omega)}|\chi_1\rangle\nonumber\\
    &=&\pm e^{i\varphi_1(\omega)}|\chi_2\rangle\nonumber\\
    &=&\pm e^{i\varphi_2(\omega)}\left(e^{i\left(\varphi_1(\omega)-\varphi_2(\omega)
    \right)}|\chi_2\rangle\right)\nonumber\\
    &\rightarrow&e^{i\left(\varphi_1(\omega)-\varphi_2(\omega)
    \right)}|\chi_2\rangle\label{a8}
    \end{eqnarray}

Again, if $\mathbf{R}$ represents a reflection (rotation) we use
the upper (lower) sign.  From these results, we see that
$\mathcal{U}_2\mathcal{U}_1$ can be treated as a single operator
$\overline{\mathcal{U}}$ with an associated phase error of
$\varphi_1(\omega)+\varphi_2(\omega)$. Factoring out the global
phase shift of $\varphi_2(\omega)$ in the last two relations
(indicated by the arrow in eqs. \ref{a7} and \ref{a8}), we see
that $\mathbf{R}\mathcal{U}_2\mathbf{R}\mathcal{U}_1$ can be
treated as a single operator
$\overline{\mathcal{U}}_{\scriptscriptstyle{QC}}$ with an
associated phase error of $\varphi_1(\omega)-\varphi_2(\omega)$.

\subsection{Experimental Measures of Coherence}\label{vis_proof}
\subsubsection{Proof of Equation \ref{visibility}}
\par We begin by defining $\lambda$ in Eq. \ref{V_lambda} to be an
equal superposition of the basis states $|\chi_j\rangle$:
\begin{subequations}
\begin{eqnarray}
|\lambda\rangle &=&
\frac{1}{\sqrt{2}}\left(|\chi_1\rangle+|\chi_2\rangle\right)\\
|\overline{\lambda}\rangle &=&
\frac{1}{\sqrt{2}}\left(|\chi_1\rangle-|\chi_2\rangle\right).
\end{eqnarray}
\end{subequations}
Then we have
\begin{eqnarray}
\mathcal{V}(x)&=&\langle\lambda|\rho(x)|\lambda\rangle-
\langle\overline{\lambda}|\rho(x)|\overline{\lambda}\rangle\nonumber\\
 &=&\frac{1}{2}\left\{(\langle\chi_1|+\langle\chi_2|)\rho(x)(|\chi_1
 \rangle+|\chi_2\rangle)\right.\nonumber\\
 &&{}-\left.(\langle\chi_1|-\langle\chi_2|)\rho(x)(|\chi_1
 \rangle-|\chi_2\rangle)\right\}\nonumber\\
&=&\left(\langle\chi_1|\rho(x)|\chi_2\rangle+\langle\chi_2|\rho(x)|\chi_1\rangle\right)\nonumber\\
&=&\left(\langle\chi_1|\rho(x)|\chi_2\rangle+\langle\chi_1|\rho(x)|\chi_2\rangle^\ast\right)\nonumber\\
&=&2Re(\langle\chi_1|\rho(x)|\chi_2\rangle)
\end{eqnarray}
where $Re$ indicates the real part.
\begin{widetext}
\subsection{Explicit Calculation of Two-Photon Density
Matrices}\label{density_matrices}

\par We begin by modeling two-photon density matrices
including frequency anti-correlation effects. Substituting the
appropriate phase functions (Eq. \ref{phase_before_modification})
into Eq. \ref{rho2}, we have an explicit expression for
 $\rho=\rho(\mathcal{L},\mathcal{L})$:
\begin{equation}\label{rho2_noDFS}
\rho=
\sum_{i,j,k,l}c_{ij}c_{kl}^{\ast}|\chi_{ij}\rangle\langle\chi_{kl}|\int\mathrm{d}
\epsilon\ \left\{ |A(\frac{\omega_o}{2}+\epsilon)|^{2}|A(\frac{\omega_o}{2}-\epsilon)|^{2} 
e^{i\left(\varphi_{i}(\frac{\omega_o}{2}+\epsilon,\mathcal{L})+
\varphi_{j}(\frac{\omega_o}{2}-\epsilon,\mathcal{L})
-\varphi_{k}(\frac{\omega_o}{2}+\epsilon,\mathcal{L})
-\varphi_{l}(\frac{\omega_o}{2}-\epsilon,\mathcal{L})\right)}\right\}.
\end{equation}
We will now calculate $\rho$ assuming the initial state matrix
elements, $c_{ij}c^\star_{kl}$, are given by the density matrix
representation of the four Bell States,
\begin{equation}
|\phi^\pm\rangle=\frac{1}{\sqrt{2}}(|HH\rangle\pm|VV\rangle),\
|\psi^\pm\rangle=\frac{1}{\sqrt{2}}(|HV\rangle\pm|VH\rangle).
\end{equation}
In this way, we see the effect of $\mathbf{U}$ on each of these
states. For brevity, let
$f(\epsilon)=|A(\frac{\omega_o}{2}+\epsilon)|^{2}
|A(\frac{\omega_o}{2}-\epsilon)|^{2}$.  Then we have,
\begin{eqnarray}\label{psiplus}
|\phi^\pm\rangle\langle\phi^\pm|=\frac{1}{2}
\begin{pmatrix}
     1&0&0&\pm1
 \cr 0&0&0&0
 \cr 0&0&0&0
 \cr \pm1&0&0&1
\end{pmatrix}
&\rightarrow&\frac{1}{2}
\begin{pmatrix}
     1 & 0 & 0 & \pm e^{-i\omega_o\frac{\mathcal{L}\Delta n}{c}}
 \cr 0 & 0 & 0 & 0
 \cr 0 & 0 & 0 & 0
 \cr \pm e^{i\omega_o\frac{\mathcal{L}\Delta n}{c}} & 0 & 0 & 1
\end{pmatrix}\\&&\nonumber \\
|\psi^\pm\rangle\langle\psi^\pm|=\frac{1}{2}
\begin{pmatrix}
     0&0&0&0
 \cr 0&1&\pm1&0
 \cr 0&\pm1&1&0
 \cr 0&0&0&0
\end{pmatrix}
&\rightarrow&\frac{1}{2}
\begin{pmatrix}
     0 & 0 & 0 & 0
 \cr 0 & 1 & \pm\int\mathrm{d}\epsilon f(\epsilon) e^{2i\epsilon\frac{\mathcal{L}\Delta n}{c}} & 0
 \cr 0 & \pm\int\mathrm{d}\epsilon f(\epsilon) e^{-2i\epsilon\frac{\mathcal{L}\Delta n}{c}} & 1 & 0
 \cr 0 & 0 & 0 & 0
\end{pmatrix}.
\end{eqnarray}

Note that the dependence on $\epsilon$ (the variable of
integration) drops out of the $(11,22)$ and $(22,11)$ terms of
$|\phi^\pm\rangle\langle\phi^\pm|$ so that there is no decoherence
due to the frequency distribution of the down-converted photons.
However, $\epsilon$ does not drop out of the $(12,21)$ and
$(21,12)$ terms of $|\psi^\pm\rangle\langle\psi^\pm|$. Due to the
finite width of the frequency term $f(\epsilon)$, for large
$\mathcal{L}$, the integral in these terms approaches zero in the
same manner as the Fourier integrals of Sect. \ref{review}.  Note
that the results developed above apply only to decoherence in the
$H/V$ basis.  The states $|\phi^\pm\rangle$ look different in
other bases, so they are subject to decoherence in general.

\par In order to model decoherence in other bases, and also to
model non-collective decoherence, we must include rotation
operators in Eq. \ref{rho2_noDFS}.  We define an operator ${\cal
R}(\theta_L,\theta_R)$ which represents a rotation by $\theta_L$
in arm L and a rotation by $\theta_R$ in path R. It can be shown
that in the present matrix notation,
\begin{equation}\label{rotation}
{\cal R}(\theta_L,\theta_R) = \begin{pmatrix}
\cos(\theta_L)\cos(\theta_R) & -\sin(\theta_L)\cos(\theta_R) &
-\cos(\theta_L)\sin(\theta_R) & \sin(\theta_L)\sin (\theta_R)\\
\sin(\theta_L)\cos(\theta_R) & \cos(\theta_L)\cos(\theta_R) &
-\sin(\theta_L)\sin(\theta_R) & -\cos(\theta_L)\sin (\theta_R)\\
\cos(\theta_L)\sin(\theta_R) & -\sin(\theta_L)\sin(\theta_R) &
\cos(\theta_L)\cos(\theta_R) & -\sin(\theta_L)\cos(\theta_R)\\
\sin(\theta_L)\sin(\theta_R) & \cos(\theta_L)\sin(\theta_R) &
\sin(\theta_L)\cos(\theta_R) & \cos(\theta_L)\cos(\theta_R)
\end{pmatrix}.
\end{equation}
It follows that ${\cal R}^{-1}(\theta_L,\theta_R)={\cal
R}^\dag(\theta_L,\theta_R)= {\cal R}(-\theta_L,-\theta_R)$.  We
can now generalize Eq. \ref{rho_formal} by writing the reduced
density operator representing polarization degrees of freedom
after an input state $\rho_\omega (0)$ has passed through crystals
of thicknesses $x_L$ and $x_R$ at angles $\theta_L$ and $\theta_R$
in the photon paths labeled L and R:
\begin{equation}\label{rho_general}
\rho(x_L,x_R,\theta_L,\theta_R)=
\iint\mathrm{d}\omega_3\mathrm{d}\omega_4\langle\omega_3|\langle\omega_4
|{\cal R}(\theta_L,\theta_R) \mathbf{U}(x_L,x_R){\cal
R}^\dag(\theta_L,\theta_R) \rho_\omega(0){\cal
R}(\theta_L,\theta_R) \mathbf{U^{\dag}}(x_L,x_R) {\cal
R}^\dag(\theta_L, \theta_R) |\omega_3\rangle|\omega_4\rangle.
\end{equation}
Eq. \ref{rho_general} can be solved numerically given
$\varphi_j(\omega,x)$ ($j\in\{1,2\}$), the eigenvalues which
define ${\bf U}(x_L,x_R)$.

\par With the same techniques, we can model density matrices
including both frequency and path anti-correlation effects.
Substituting the modified phase functions
(\ref{phase_after_modification}) into Eq. \ref{rho2_modified},
 we have an expression which includes both frequency and path
 anti-correlation effects:
\begin{equation}\label{rho2_DFS}
\rho=
\sum_{i,j,k,l}c_{ij}c_{kl}^{\ast}|\chi_{ij}\rangle\langle\chi_{kl}|\int\mathrm{d}
\epsilon\ \left\{
|A(\frac{\omega_o}{2}+\epsilon)|^{2}|A(\frac{\omega_o}{2}-\epsilon)|^{2}
e^{i\left(\tilde{\varphi_{i}}
(\frac{\omega_o}{2}+\epsilon,\mathcal{L}) + \varphi_{j}
(\frac{\omega_o}{2} - \epsilon,\mathcal{L}) -\tilde{\varphi_{k}}
(\frac{\omega_o}{2}+\epsilon,\mathcal{L})
-\varphi_{l}(\frac{\omega_o}{2}-\epsilon,\mathcal{L})\right)}\right\}.
\end{equation}
Again, using the four Bell States as input states, we see the
effect of $\tilde{\mathbf{U}}$ on each of these states.
\begin{eqnarray}
|\phi^\pm\rangle\langle\phi^\pm|=\frac{1}{2}
\begin{pmatrix}
     1&0&0&\pm1
 \cr 0&0&0&0
 \cr 0&0&0&0
 \cr \pm1&0&0&1
\end{pmatrix}
&\rightarrow&\frac{1}{2}
\begin{pmatrix}
     1 & 0 & 0 & \pm\int\mathrm{d}\epsilon f(\epsilon) e^{2i\epsilon\frac{\mathcal{L}\Delta n}{c}}
 \cr 0 & 0 & 0 & 0
 \cr 0 & 0 & 0 & 0
 \cr \pm\int\mathrm{d}\epsilon f(\epsilon) e^{-2i\epsilon\frac{\mathcal{L}\Delta n}{c}} & 0 & 0 & 1
\end{pmatrix}\\&&\nonumber \\
|\psi^\pm\rangle\langle\psi^\pm|=\frac{1}{2}
\begin{pmatrix}
     0&0&0&0
 \cr 0&1&\pm1&0
 \cr 0&\pm1&1&0
 \cr 0&0&0&0
\end{pmatrix}
&\rightarrow&\frac{1}{2}
\begin{pmatrix}
     0 & 0 & 0 & 0
 \cr 0 & 1 & \pm e^{-i\omega_o\frac{\mathcal{L}\Delta n}{c}} & 0
 \cr 0 & \pm e^{i\omega_o\frac{\mathcal{L}\Delta n}{c}} & 1 & 0
 \cr 0 & 0 & 0 & 0
\end{pmatrix}.
\end{eqnarray}

\end{widetext}
Note that in this case, the dependence on $\epsilon$ (the variable
of integration) drops out of the $(21,12)$ and $(12,21)$ terms of
$|\psi^\pm\rangle\langle\psi^\pm|$.  On the other hand, $\epsilon$
does not drop out of the $(11,22)$ and $(22,11)$ terms of
$|\phi^\pm\rangle\langle\phi^\pm|$.  Again, the results developed
above apply only to decoherence in the $H/V$ basis.
The analytical form of $\rho$ is helpful in developing a physical
picture of the decoherence process.  In order to calculate the
corresponding result for decoherence in an arbitrary basis, we
simply set $\theta_L=\theta_R-\frac{\pi}{2}$ in Eq.
\ref{rho_general} and solve numerically.  This corresponds to the
path anti-correlation procedure described earlier. These results
will be presented elsewhere \cite{DFS_paper}, however, the
theoretical predictions are unchanged for the state
$|\psi^-\rangle$, since it looks the same in every basis.

\section*{Acknowledgements}

 Experimental and theoretical work was performed in the Quantum Information
laboratory of P. G. Kwiat at LANL. We wish to thank A. G. White
for his willingness to assist in all stages of this work. At
Dartmouth, AJB gratefully acknowledges W. Lawrence and M. Mycek
for offering support, advice and feedback at various stages.

\end{document}